\documentclass[twocolumn]{aastex1}
\usepackage{soul}
\usepackage{lineno}

\usepackage{CJK}
\usepackage{graphicx}	
\usepackage{amsmath}	
\usepackage{listings}
\usepackage{acronym}
\usepackage{color}
\definecolor{dkgreen}{rgb}{0,0.6,0}
\definecolor{gray}{rgb}{0.5,0.5,0.5}
\definecolor{mauve}{rgb}{0.58,0,0.82}
\definecolor{golden}{rgb}{0.86,0.65,0.01}
\received{XXX}
\revised{YYY}
\accepted{ZZZ}

\submitjournal{ApJS}

\shorttitle{Search for compact WD binary candidates}
\shortauthors{Ren et al.}
\graphicspath{{./}{figures/}}

\begin{document}

\begin{CJK*}{UTF8}{gbsn}
\title{A systematic search for short-period close white dwarf binary candidates based on the \emph{Gaia EDR3} catalog and the \emph{Zwicky Transient Facility} data}

\correspondingauthor{Yi-Ming Hu}
\email{huyiming@mail.sysu.edu.cn}

\newcommand{\TRC}{
MOE Key Laboratory of TianQin Mission, TianQin Research Center for Gravitational Physics, Frontiers Science Center for TianQin, Gravitational Wave Research Center of CNSA, Sun Yat-sen University, Zhuhai 519082, China
}
\newcommand{\SPA}{
School of Physics and Astronomy, Sun Yat-sen University, Zhuhai 519082, China
}
\newcommand{\CSST}{
CSST Science Center for the Guangdong-Hongkong-Macau Greater Bay Area, Sun Yat-sen University, Zhuhai 519082, China
}

\acrodef{CWDB}[CWDB]{Close White Dwarf Binary}
\acrodefplural{CWDB}[CWDBs]{Close White Dwarf Binaries}
\acrodef{PCEB}[PCEB]{Post-Common Envelope Binary}
\acrodefplural{PCEB}[PCEBs]{Post-Common Envelope Binaries}
\acrodef{CV}[CV]{Cataclysmic Variable}
\acrodef{DWD}[DWD]{Double White Dwarf}
\acrodef{CRTS}[CRTS]{the Catalina Real-time Transient Survey}
\acrodef{Gaia}[Gaia]{Global Astrometric Interferometer for Astrophysics}
\acrodef{OGLE}[OGLE]{the Optical Gravitational Lensing Experiment}
\acrodef{PTF}[PTF]{the Palomar Transient Factory}
\acrodef{ASAS-SN}[ASAS-SN]{the All Sky Automated Survey for SuperNovae}
\acrodef{TESS}[TESS]{the Transit Exoplanet Survey Satellite}
\acrodef{ZTF}[ZTF]{the Zwicky Transient Facility}
\acrodef{LAMOST}[LAMOST]{the Large Sky Area Multi-Object Fiber Spectroscopic Telescope}
\acrodef{MS}[MS]{main sequence}
\acrodef{WDCS}[WDCS]{white dwarf cooling sequence}
\acrodef{SNR}[SNR]{signal-to-noise ratio}
\acrodef{RMS}[RMS]{root mean square}
\acrodef{spCV}[spCV]{short-period cataclysmic variable}
\acrodef{LS}[LS]{Lomb-Scargle}
\acrodef{CE}[CE]{Conditional Entropy}
\acrodef{BLS}[BLS]{Box Least Squares}
\acrodef{GW}{gravitational wave}

\author[0000-0002-1428-4003]{Liangliang Ren (任亮亮)}
\affiliation{\SPA}
\affiliation{\TRC}

\author[0000-0002-3084-5157]{Chengyuan Li (李程远)}
\affiliation{\SPA}
\affiliation{\CSST}

\author{Bo Ma (马波)}
\affiliation{\SPA}
\affiliation{\CSST}

\author[0000-0002-9156-7461]{Sihao Cheng (程思浩)}
\affiliation{Department of Physics and Astronomy, The Johns Hopkins University, 3400 N Charles Street, Baltimore, MD 21218, USA}
\affiliation{Centre de Sciences des Donn\'ees, Ecole Normale Sup\'erieure, 45 Rue d'Ulm, 75005, Paris, France}

\author[0000-0002-7112-759X]{Shun-Jia Huang (黄顺佳)}
\affiliation{\SPA}
\affiliation{\TRC}

\author[0000-0002-0066-0346]{Baitian Tang (汤柏添)}
\affiliation{\SPA}
\affiliation{\CSST}

\author[0000-0002-7869-0174]{Yi-ming Hu (胡一鸣)}
\affiliation{\SPA}
\affiliation{\TRC}



\begin{abstract}

Galactic short-period close white dwarf binaries (CWDBs) are important objects for space-borne gravitational-wave (GW) detectors in the millihertz frequency bands. Due to the intrinsically low luminosity, only about 25 identified CWDBs are detectable by LISA, which are also known as verification binaries. The Gaia Early Data Release 3 (EDR3) provids a catalog containing a large number of CWDB candidates, which also includes \texttt{parallax} and photometry measurements. We cross-match the Gaia EDR3 and Zwicky Transient Facility public data release 8, and apply period-finding algorithms to obtain a sample of periodic variables. The phase-folded light curves are inspected, and finally we obtain a binary sample containing 429 CWDB candidates. We further classify the samples into eclipsing binaries (including 58 HW Vir-type binaries, 65 EA-type binaries, 56 EB-type binaries, 41 EW-type binaries) and ellipsoidal variations (209 ELL-type binaries). We discovered 4 ultra-short period binary candidates with unique light curve shapes. We estimate the GW amplitude of all our binary candidates, and calculate the corresponding signal-to-noise ratio for TianQin and LISA. We find 2 (6) potential GW candidates with signal-to-noise ratio greater than 5 in the nominal mission time of TianQin (LISA), which increases the total number of candidate verification binaries for TianQin (LISA) to 18 (31).

\end{abstract}



\section{Introduction} \label{sec:intro}

\acp{CWDB} are an important branch of the evolution channel of main-sequence star binaries \citep{Gokhale2007}.
The formation of such binaries involves a stage of the common envelope, a complicated stage where complicated physical processes such as mass transfer, angular momentum loss, and even gravitational wave radiation.
The theoretical simulation of the \acp{CWDB} evolution is challenging, with major issues still under debate. 
For example, it is not clear whether a second common envelope stage is involved in the formation of a \ac{CWDB}.
Since there are many sub-classes of \acp{CWDB}, there could be a variety of evolution channels.
The searching and identification of new close white dwarf binaries have the potential to provide observational evidence for the binary evolution models and are important for the research of stellar physics, milli-Hertz gravitational-wave astronomy, and Galactic evolution.\citep{Parsons2016,Rebassa2017,Lagos2020,Ren2020,Hernandez2021,Hernandez2022,Lagos2022}. 

Depending on the components type and mass transfer, \acp{CWDB} can be divided into several sub-populations \citep{Ren2020,Inight2021,Kruckow2021}: \acp{PCEB}, binary systems that contains a white dwarf and a main-sequence star; \acp{CV}, interactive white dwarf binaries; and \acp{DWD}. 

Short-period \acp{CWDB} have orbital periods of less than 60 minutes, emitting \ac{GW} in the millihertz-frequency band, becoming ideal sources for the space-borne gravitational wave observatories like the Laser Interferometer Space Antenna (\emph{LISA}) \citep{Amaro-Seoane2017} and \emph{TianQin} \citep{Luo2016,Gong2021}, therefore, they are also known as verification binaries (VB)\citep{Kupfer2018,Huang2020}. 
The known classical verification binaries catalog includes 11 AM CVn-type binaries, 13 detached double white dwarfs, and 1 hot subdwarf binaries \citep{Kupfer2018}. 
Up to now, a total of $\sim$ 87 candidate verification binaries are all found in the electromagnetic band \citep{Huang2020,Burdge2019a,Burdge2020b,Burdge2020a,Kilic2021,Chandra2021,Brown2022}.
In addition, there are other potential \ac{GW} sources such as Helium star binaries, short-period CVs, exoplanets, brown dwarfs, and substellar companies \citep{Amaro-Seoane2022}. 
From the perspective of the formation pathways, these verification binaries and other potential \ac{GW} sources experienced mass exchange phases via Roche-lobe overflow and mass loss via the ejection of a common envelope (CE) in evolution \citep{Gotberg2020,Amaro-Seoane2022}. For example, there may be two different channels for the formation of DWDs: if the mass transfer stage is stable, it will be directly generated from CVs after a CE phase. If the mass transfer is unstable, the second CE stage takes place after the CVs, and then it will evolve into a WD+Helium-star, and finally form a DWD \citep{Amaro-Seoane2022}. It is noteworthy that \citet{El-Badry2021b} recently found evidence that the progenitors of extremely low mass white dwarfs (ELM WDs), AM CVn systems, and detached ultracompact binaries may be evolved CVs \citep{El-Badry2021b,El-Badry2021a}. The Galactic foreground is mainly composed of DWDs, but other Galactic binaries also contribute to the Galactic foreground \citep{Boileau2021}, for example, CVs, stripped star binaries \citep{Gotberg2020}, WD+M-dwarfs, etc.

In addition to searches from individual observations, one can perform \ac{CWDB} search with a larger scale.
Several surveys can be used to pick potential \ac{CWDB} candidates, like 
\ac{CRTS} \citep{Breedt2014}, 
\ac{Gaia} \citep{Geier2017,Gentile2021,Rebassa2021,El-Badry2021c,Torres2022}, 
\ac{OGLE} \citep{Wevers2016}, 
\ac{PTF} \citep{vanRoestel2017,vanRoestel2018,Burdge2019b}, 
\ac{ASAS-SN} \citep{RiveraSandoval2022}, 
\ac{TESS} \citep{WangKun2020,PichardoMarcano2021,Hernandez2022},
\ac{LAMOST} \citep{WangKun2022},
and most recently \ac{ZTF} \citep{Ofek2020,Coughlin2020,Burdge2020b,Burdge2020a,Szkody2020,Szkody2021,Kupfer2021,Keller2022,McWhirter2022}.
The repeated observations of the same sources can reveal the optical variability, and some of these surveys can be used, and have been used, to determine the periodicity.

In this paper, we present a search for \acp{CWDB} from a combination of survey data.
We first select potential variable sources from the Gaia EDR3 catalog, then apply periodic analysis from time domain photometric data from \ac{ZTF} data to determine the properties of the binaries.

The structure of this article is organized as follows. In Section~\ref{sec:closewdbs}, we elaborate the classification and the characteristics of the different sub-classes of close white dwarf binaries and the status quo of potential gravitational wave sources for these subtypes. In Section~\ref{sec:gaiasample}, we describe how to select our initial samples based on the Gaia H-R diagram, and further select the variable sample by the Gaia Variability Metric. In Section~\ref{sec:ztfpdr8}, we describe how we use three period-finding algorithms to search for periodic samples from ZTF light curves. In Section~\ref{sec:analaysisresults}, we analyze five different variability types of binaries in our binary sample and compare our sample with the CWDBs catalog of other surveys. In Section~\ref{sec:discussion}, we discuss four interesting binaries. The overall parameter distribution and \ac{GW} properties of the final sample are also discussed. Finally we summarize in Section~\ref{sec:conclusion}.

\section{Close White Dwarf Binaries and potential gravitational wave sources} \label{sec:closewdbs}

There are two evolutionary pathways for primordial binaries: 
(1) The so-called wide binary, about 75\% of wide binaries have relatively large orbital separations, and there is no interaction in the Hubble time-scale, which is similar to the evolution of a single star \citep{Ren2020}. 
(2) The remaining 25\% are close binaries, which have to undergo a phase of mass transfer \citep{Ren2020}.
After the ejection of the common-envelope (CE) phase, at least one star in the close binary evolved into a white dwarf, and due to complex physical processes such as angular momentum loss or gravitational wave radiation, which will be generated sub-classes such as \acp{PCEB}, \acp{CV}, and \acp{DWD}. 

In this paper, we will focus on the Galactic ultra-compact white dwarf binary candidates that can generate gravitational wave radiation, potential gravitational wave sources, and the progenitors of \ac{GW} sources. 
These targets are formed by the evolutionary pathway of the primordial binaries after the CE stage \citep{Kruckow2021}. 
We will focus on the classification and the characteristics of each sub-type of \acp{CWDB}: \acp{PCEB}, \acp{CV}, and \acp{DWD}, respectively.

\subsection{Post-Common Envelope Binaries}

PCEBs are detached or semi-detached binary systems in which a white dwarf and main-sequence star that is rapidly engulfed in a common envelope, and the orbital periods range from hours to a day. The companion stars in the PCEBs system are low-mass main-sequence stars, such as several brown dwarfs, M-dwarfs, hot subdwarfs, etc. Brown dwarf companion stars are difficult to search from optical band data alone, while M-dwarf and hot subdwarf companions are in the majority. Hot subdwarfs are under-luminous for their high temperature\citep{Heber2016,Lei2020,Culpan2022}. They are divided into two main spectroscopic: B-type subdwarfs (sdB, a core helium-burning star and thin hydrogen envelope), and O-type subdwarfs (sdO). Hot subdwarf binaries with white dwarf companions that exit the common envelope phase at orbital periods of less than two hours will overflow their Roche lobes while the sdB is still burning helium\citep{Kramer2020,Li2022}.

Striped star binaries are one of the populations of \ac{GW} sources that can be detected by \emph{LISA} \citep{Gotberg2020}. Among them, the short-period hot subdwarf binaries are potential verification binaries for generating \ac{GW} radiation. Until recently, only six known the ultra-compact sdB+WD-type binaries are VB sources with orbital periods less than 100 min, namely CD-30$^{\circ}$11223 \citep{Geier2013, Kupfer2018}, HD 265435 \citep[TIC 68495594,][]{Pelisoli2021}, ZTF J1946+3203, ZTF J0640+1738, ZTF J2130+4420, ZTF J2055+4651 \citep{Burdge2020a, Kupfer2020a, Kupfer2020b}.

\begin{figure*}[htpb]
  \begin{center}
  \includegraphics[width=0.98\textwidth]{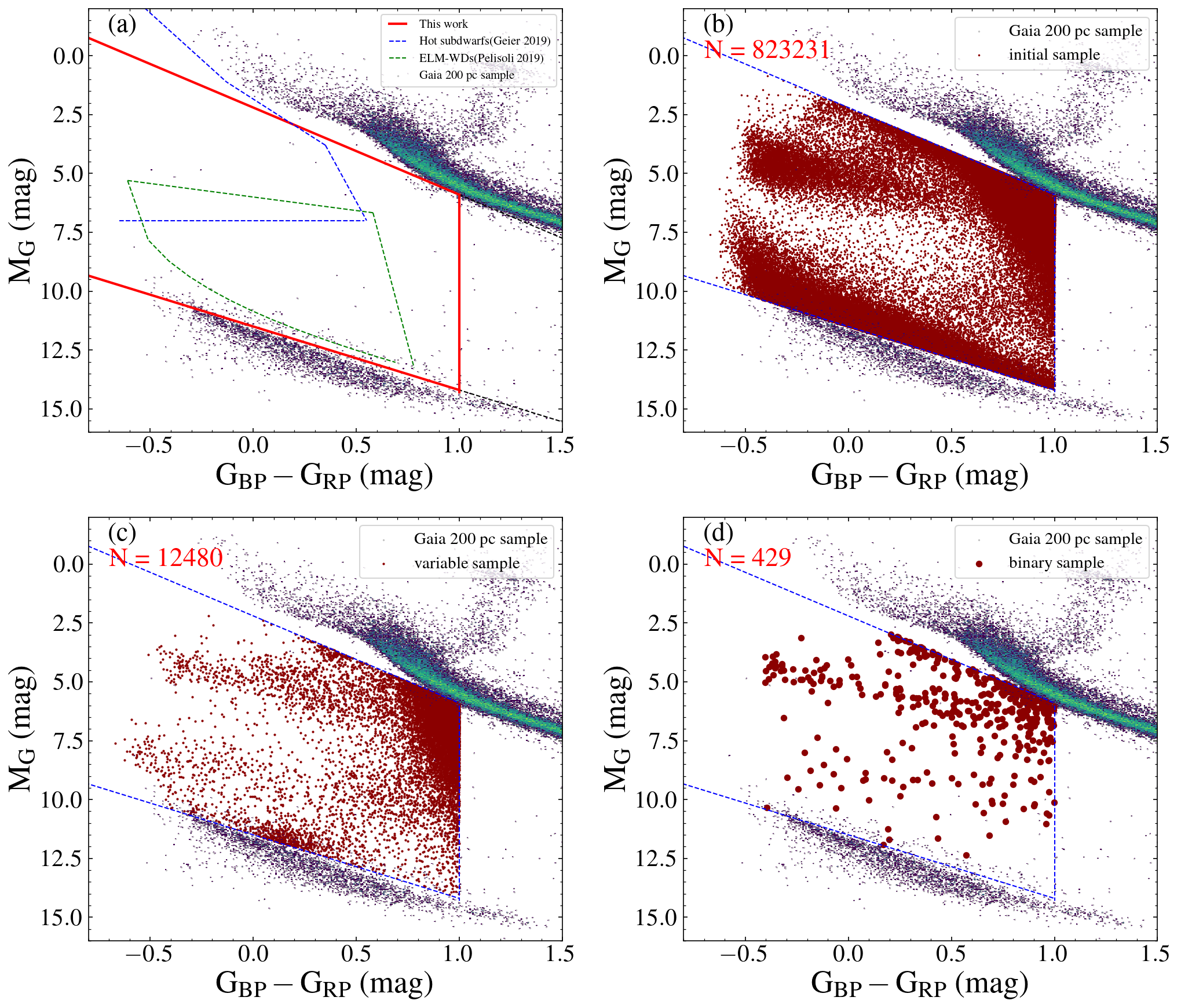}
  \caption{The location of the sample selection on the Gaia color-magnitude diagram. In all panels, scatter points and the overlaid 2D histogram shows the stellar density of the Gaia 200 pc background sources. (a) The red solid lines show the color-cut range that we defined, using the lower edge of the main sequence as an upper limit and the upper edge of the WD cooling sequence as a bottom limit, and using a color-cut \texttt{$\rm G_{BP} - G_{RP}$} $<$ 1.0 as a right color limit. The blue dashed lines show a color-cut and absolute magnitude selection scheme for searching hot subdwarfs using Gaia data proposed by \citet{Geier2019}. The green dashed lines show the color-magnitude diagram selection scheme for searching ELM-WD candidates based on Gaia DR2 data by \citet{Pelisoli2019b}. (b) The red scatter points shows all samples under the initial query conditions. (c) The red scatter points shows the variable samples after using the Gaia variability metric select condition. (d) The red scatter points shows the final binary samples after examining ZTF light curve data.}
  \label{fig:GaiaSampleHR}
  \end{center}
\end{figure*}

\subsection{Cataclysmic Variables}

Cataclysmic variables are short-period ($\mathrm{P}_{\rm orb}\leq$ 6 h) semi-detached binary star systems in which a low-mass companion star transfers mass to a white dwarf (WD) primary via stable Roche lobe overflow\citep[RLOF,][]{Knigge2011,Hillman2020,Schreiber2021,Schreiber2022}. The two stars are close enough that the companion completely fills its Roche lobe, and the outer layer of its envelope is gradually stripped from its surface and forms an accretion disk around the white dwarf. The CVs can be further divided into several sub-types: classical novae, dwarf novae, polar \citep[AM Her-type star,][]{Kolbin2020,Kolbin2022}, intermediate polar \citep[DQ Her star,][]{Norton1999} and AM Canum Venaticorum system\citep[AM CVn,][]{Nelemans2001,Postnov2014,Ramsay2018,Wong2021}, etc. 

For the polar (B $\sim$10-200 MG) and intermediate polar (B $\sim$1-10 MG), the primary is a highly magnetized WD, and the accretion disk can be entirely disrupted. There are differences between polar and intermediate polars, the rotation of the white dwarf is synchronized with the orbital period of the companion in the polars, while the latter is not synchronized. 

The AM CVn systems are short-period ultra-compact binaries in which a massive white dwarf accretes helium-rich matter from a hydrogen-deficient donor star. For AM CVn systems, the donors star are degenerate or semi-degenerate, and the orbital periods are in the range of $\simeq$ 5-68 minutes\citep{Kalomeni2016,Green2018,Green2020,vanRoestel2021b,vanRoestel2021a,Duffy2021}. 

\subsection{Double White Dwarfs}

Double white dwarfs are the remnants of the second common envelope event in the evolution of close binaries. DWDs lose angular momentum through gravitational wave radiation, which will decrease the separation between the two white dwarfs and may eventually merge. Several relics can be associated with the double white dwarf merging event, such as type Ia supernovae \citep{Nelemans2001a,Toonen2012,Shen2018,Liu2018,Rebassa2019,Maselli2020}, neutron stars \citep[Accretion-Induced Collapse events,][]{Yu2019,Liu2020,Ablimit2022,Wang2022} and high-mass WDs \citep{Cheng2020}. The exact outcome depends on the type of core of the two white dwarfs. 

Extremely low-mass WDs are double-degenerate helium-core WDs with masses $M\leq 0.3 M_{\odot}$. ELM WD systems are formed after severe mass loss, which limits that these systems most likely formed through binary interactions\citep{Li2019}. Because it is difficult to evolve into ELM WD from an isolated star in the Hubble timescale. The ELM survey is systematic search research for ELM WD binaries. This survey selects candidates by using SDSS color-cutting, then carried out the targeted spectroscopic follow-up of ELM WD candidates. The ELM survey successfully discovered 124 detached double-white dwarfs (DWDs)\citep{Brown2010,Kilic2011,Brown2012,Kilic2012,Brown2013,Gianninas2015,Brown2016,Brown2020a,Kosakowski2020,Brown2022}, including eleven systems which are strong candidate \emph{LISA}-detectable gravitational-wave sources \citep{Brown2011b, Kilic2014, Brown2020b, Brown2022}.

\section{Target Selecion from Gaia EDR3} \label{sec:gaiasample}

\subsection{Gaia Color-Magnitude Diagram Selection} \label{subsec:colorselection}

Gaia Early Data Release 3 \citep[Gaia EDR3,][]{Lindegren2021} contains more than 1.8 billion sources (the magnitude range from $\mathrm G = 3 - 21$) with astrometric and photometric data based on the first 34 months of observations by the Gaia satellite. The astrometric data in the Gaia EDR3 provides positions, proper motions, and parallaxes ($\varpi$) for 1.468 billion sources, and the photometric data includes 1.544 billion sources which have magnitudes of photometric three-passbands G ($\texttt{phot\_g\_mean\_mag}$), $\rm G_{BP}$, and $\rm G_{RP}$ (the BP and RP are defined by the blue and red photometers).

Thanks to the excellent astrometry measurement ability of Gaia, it can provide an accurate estimate of stars' parallax. 
Compared with the Gaia DR2, the accuracy of parallax measurements of the Gaia EDR3 is improved by 20-30 percent on average.
Meanwhile, the photometric observation ability enables the Gaia EDR3 to also provide color \texttt{$\rm G_{BP} - G_{RP}$}, and the G-band apparent magnitude of stars.
Therefore, we can use the color-magnitude diagram to select our sample of object sources.
Based on the distribution characteristics of known \ac{CWDB} systems in the H-R diagram \citep{Pelisoli2019b,Inight2021}, we find that the absolute magnitudes of these systems are dimmer than \ac{MS} stars and brighter than single white dwarfs. 
We decide to apply the first cut in the color-magnitude diagram between the \ac{MS} and \ac{WDCS}.
We further apply a selection condition on color \texttt{$\rm G_{BP} - G_{RP}$} $<$ 1.0 to favor bluer objects, as white dwarfs are expected to have high temperatures.
These first selection conditions are illustrated in Figure~\ref{fig:GaiaSampleHR}(a) as red solid lines. 
The corresponding equations are as follows:

\begin{equation}
      \label{eq:equation1}
      {\rm G_{BP} - G_{RP}} < 1.0,
\end{equation}

\begin{equation}
      \label{eq:equation2}
      {\rm G_{abs}  <  11.5 + 2.7\left(G_{BP} - G_{RP}\right)},
\end{equation}

\begin{equation}
      \label{eq:equation3}
      {\rm G_{abs}  >  2.2 + 3.7\left(G_{BP} - G_{RP}\right)},
\end{equation}
where $\rm G_{abs} = G + 5*log10\left[(\varpi + 0.029)/1000\right] + 5$ is the absolute magnitude \citep{Pelisoli2019b}. Our cutting scheme includes the selection range of extremely low-mass white dwarf candidates (green dashed lines in Figure~\ref{fig:GaiaSampleHR}(a)), and includes hot subdwarf stars \citep{Geier2019} (blue dashed lines in Figure~\ref{fig:GaiaSampleHR}(a)).

To ensure that the red and blue photometry are not subject to random noise, we follow \citep{Pelisoli2019a,Pelisoli2019b} and filtered on the errors of both $\rm G_{BP}$ and $\rm G_{RP}$ flux measurements to be larger than 10 (\texttt{phot\_bp/rp\_mean\_flux\_over\_error > 10}). 
We also apply a \ac{SNR} threshold of 5 on the parallax measurements (\texttt{parallax\_over\_error  > 5}).
The filter conditions for the threshold of \texttt{parallax\_over\_error  > 5} are often adopted for searches with Gaia data, for example, the search for ELM-WD candidates \citep{Pelisoli2019b}, the search for pulsating white dwarfs \citep{Guidry2021}, the search for evolved-CVs \citep{El-Badry2021b}, and the search for AR Scorpii-type binary systems \citep{Takata2022}. 
Since the motivation of this work is to obtain reliable candidate \ac{GW} sources, we conclude that precise measurement of parallaxes is very important so that the estimation of \ac{GW} SNR can be trustworthy.
By applying this filter, we exclude a large number of sources with positive and negative spurious parallaxes, especially near the color \texttt{BP-RP} $\sim 1$ in the Gaia H-R diagram.

\citet{Lindegren2018,Lindegren2021} proposed to apply the quality filter parameters selection condition on Gaia DR2, and Pelisoli \citep{Pelisoli2019a,Pelisoli2019b} used similar conditions to the study of ELM-WD candidates.
In this work, we apply the flux excess factor \texttt{phot\_bp\_rp\_excess\_factor}, $E$ as the quality filter parameters to the selection:
\begin{equation}
      \label{eq:equation4}
      E < 1.45 + 0.06\left(G_{BP} - G_{RP}\right)^{2},
\end{equation}
\begin{equation}
      \label{eq:equation5}
      E > 1.0 + 0.015\left(G_{BP} - G_{RP}\right)^{2},
\end{equation}
\begin{equation}
      \label{eq:equation6}
      u < 1.2\max(1, \exp(-0.2(G - 19.5))),
\end{equation}
where $\rm G$ is the G-band mean magnitude (\texttt{phot\_g\_mean\_mag}), and u is defined by parameters the value of the chi-square statistic of the astrometric solution, \texttt{astrometric\_chi2\_al}, and the number of good observations, \texttt{astrometric\_n\_good\_obs\_al}:
\begin{equation}
      \label{eq:equation6}
      u = \sqrt{\mathrm {astrometric\_chi2\_al / (astrometric\_n\_good\_obs\_al - 5)}},
\end{equation}

We plan to search for periodicity from the \ac{ZTF} time-domain photometric data.
However, the geographical site of \ac{ZTF} put a limit on the archived data, so we apply a final cut in declination of $\delta > -28$ deg \citep{Bellm2019,Graham2019,Masci2019,Dekany2020}. For further details on the initial-cut selection criteria, please see Appendix \ref{ref:appendixa}.

\begin{figure}
  \centering
  \includegraphics[width=\columnwidth]{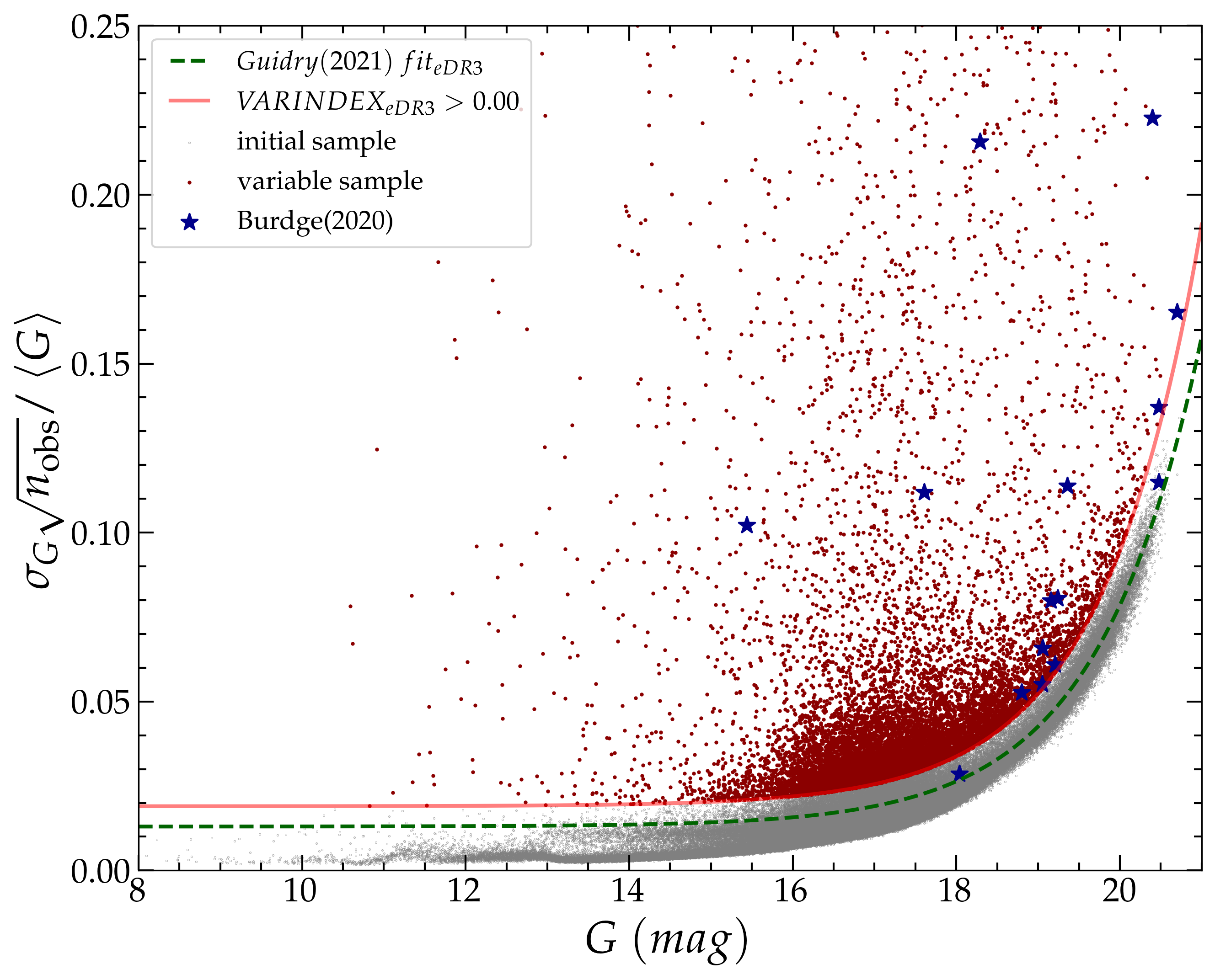}
  \caption{Empirical Gaia variability metric as a function of G-band magnitude for all initial samples, with the exponential fit proposed by \citet{Guidry2021} for Gaia EDR3 delineated in green. The 1.5 percent most variable samples are marked in red, and 15 ultra-compact binaries reported by \citet{Burdge2020b} are marked as blue stars.}
  \label{fig:Metric}
\end{figure}

\begin{figure}
  \centering
  \includegraphics[width=\columnwidth]{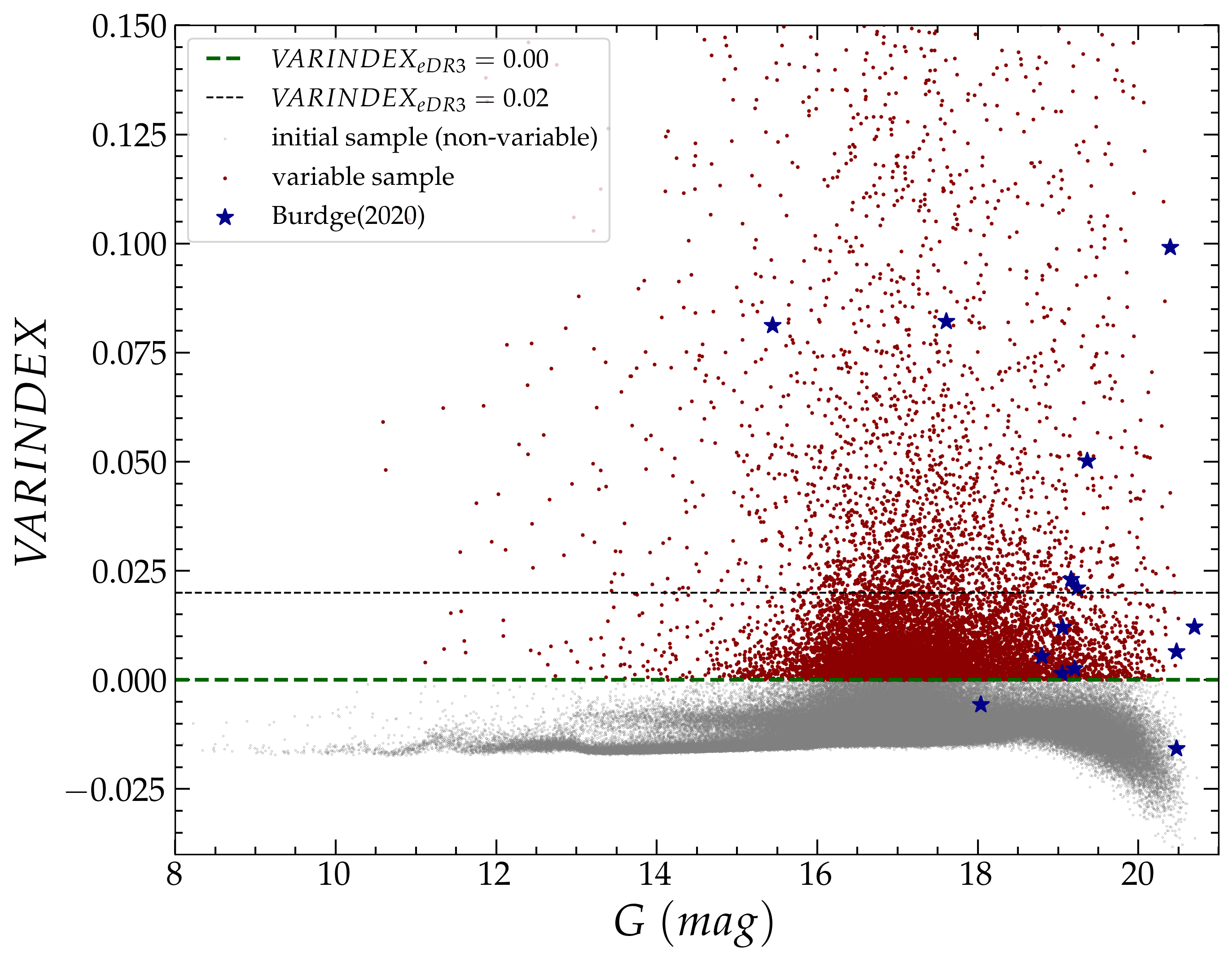}
  \caption{The $\texttt{VARINDEX}$ and G band magnitude of the 12,480 candidates after Gaia variability cut. The green dashed line shows that 1.5 \% of most variable samples have VARINDEX $>$ 0.00. The grey dashed line show \citet{El-Badry2021b} used VARIANDEX $>$ 0.02 to cut the samples in search of evolved-\acp{CV}. The 1.5 percent most variable samples are marked in red, and 15 ultra-compact binaries reported by \citet{Burdge2020b} are marked as blue stars.}
  \label{fig:VARINDEX}
\end{figure}

\subsection{Gaia Variability Metric}

In the above section, in order to obtain clean initial samples, we have used a series of selection conditions, such as the the error of parallax measurements, and the quality of filtering parameter.
Now our samples are accompanied with precise distance measurement and reliable photometric data. 
In this step, we aim to identify the variable sources that are most likely to be  binaries.
Therefore, we apply a further selection to the candidates from another dimension: photometric variability.
The period search relies on the change of luminosity over time, and we can discard candidates with stable magnitudes, please see Appendix \ref{ref:appendixa}.
We apply a cut in the Gaia variability metric to select variable sources in the initial sample.
We adopt the ``variable amplitude" proposed by \citet{Deason2017} based on the \ac{RMS} dispersion of flux.
A number of studies have adopted the Gaia variable metric and successfuly identified different types of variable sources, including large-amplitude variables \citep{Mowlavi2021}, \ac{spCV} \citep{Abrahams2020}, pulsing white dwarf \citep{Guidry2021},  hot subdwarf stars \citep{Barlow2022}, and birth of the extremely low mass white dwarfs \citep{El-Badry2021b, El-Badry2021a}.
Following \citet{Guidry2021} and \citet{El-Badry2021b}, we use the G-band photometris to further filter candidate with Gaia variability metric (or the variability amplitude proxy $\mathrm A_{\rm proxy,G}$, see \citet{Mowlavi2021}), $\rm V_{G}$:
\begin{equation}
      \label{eq:equation7}
      V_{G}=\frac{\sigma_{G}}{\left\langle G\right\rangle }\sqrt{n_{{\rm obs}}},
\end{equation}
where $n_{\rm obs}$ is the number of observations contributing to the G photometry (\texttt{phot\_g\_n\_obs}), $\sigma_{G}$ is the error on G-band mean flux (\texttt{phot\_g\_mean\_flux\_error	
}), $\left\langle \rm G\right\rangle$ is the G-band mean flux  (\texttt{phot\_g\_mean\_flux}).
Figure~\ref{fig:Metric} shows the Gaia variability metric distribution of the 823231 objects in the initial sample.

One can notice from Figure~\ref{fig:Metric} that the Gaia variability metric is actually dependent on the G-band magnitude.
This is because for dimmer objects, the associated \ac{SNR} will be smaller, leading to a larger contribution from random fluctuation.
In order to select the most-probable variable sources from the initial sample, we follow \citet{Guidry2021} and define the \texttt{VARINDEX} to cut the top 1\% most variable \acp{CWDB} for Gaia EDR3:
\begin{equation}
      \label{eq:equation8}
      \texttt{VARINDEX} = V_{G} - (Ae^{\alpha G} + Be^{G - 17.0} + C),
\end{equation}
where A = $8.31\times 10^{-9}$, $\alpha = 0.794$, B = 0.0005, and C = 0.00962.
In Figure~\ref{fig:VARINDEX}, we show the distribution of the \texttt{VARINDEX} over G-band magnitude.
We obtained the {\emph{variable sample}}, which contains the 12480 strongest candidate variables with $\texttt{VARINDEX} > 0.00$, as shown in Figure~\ref{fig:GaiaSampleHR}(c).
It corresponds roughly to the 1.5\% highest \texttt{VARINDEX} systems. We conclude that this choice is reasonable, since if we list the \texttt{VARINDEX} values of the 15 ultra-compact binaries reported by \citet{Burdge2020b}, 13 of them are associated with $\texttt{VARINDEX} > 0.00$,  as shown in Figure \ref{fig:VARINDEX}. Even though 2 sources are discarded in this stage, this selection retains most of the interesting sources while greatly shrinking the sample size.

\subsection{Variable Sample}

We used the Gaia Variability flags to select 12480 candidates (see Table \ref{tab:sampletype}), which is defined as a {\emph{variable sample}} of objects for period search. 
We imposed a photometric color selection of \texttt{${\rm G_{BP} - G_{RP} < 1.0}$}, this is different from the color-cut scheme used by \citet{Pelisoli2019b} to search for ELM candidates. \citet{Pelisoli2019b} used a color-cut criteron of \texttt{${\rm G_{abs} > 20.25(G_{BP} - G_{RP}) - 7.15}$} to search for blue targets to reduce contamination sources such as faint red stars scattered from the main sequence.
We are indeed interested in ``blue" objects for the Gaia source catalog because we expected short-period CWDBs have higher temperatures due to tidal heating \citep{Burdge2020a}, especially those candidates whose orbital period is less than 60 min may become strong \ac{GW} sources. However, we are also interested in other potential gravitational wave source candidates, such as short-period CVs and substellar companies \citep{Amaro-Seoane2022}. 
It is worth noting that our {\emph{variable sample}} includes some outbursting objects, such as nonmagnetic CV (dwarf nova), magnetic CV (irregular variability on a wide range of timescales), and QSO (red-shifted emission lines) \citep{Pelisoli2019b}, but these sources are considered as contamination in this study.

\section{Zwicky Transient Facility Photometry} \label{sec:ztfpdr8}
\subsection{ZTF Light Curves}

The \ac{ZTF} is a time-domain survey using the 48-inch (P48) Schmidt telescope at Palomar Observatory\footnote{\url{http://www.ztf.caltech.edu}}.
Its camera has a field of view of 47 square degrees, and it can scan the entire northern visible sky (Southern sky $\rm dec > -28$ deg.) at a rate of $\sim$ 3750 $\rm deg^{2}$ per hour \citep{Bellm2019,Graham2019,Masci2019,Dekany2020}.
\ac{ZTF} can reach a 5$\sigma$ limiting apparent magnitude of $\sim$ 20.8 mag in the $g$ band, $\sim$ 20.6 mag in the $r$ band, and $\sim$ 20.2 mag in the $i$ band in a 30s exposures.

For the selected variable sample, we cross match the coordinate with the \ac{ZTF} data, by downloading light curve data from the ZTF Public Data Release 8 (ZTF DR8)\footnote{\url{https://www.ipac.caltech.edu}}$^{,}$\footnote{\url{https://www.ztf.caltech.edu/page/dr8}}.

\subsection{Period Finding}

The 12480 variable samples are not necessarily binary systems.
To verify the binary nature of the system, the identification of the light curve period is necessary.
To do so, we first cross-match the variable sample from Gaia data with the public \ac{ZTF} 8th release data and identified 10938 sources that are accompanied by \ac{ZTF} light curves.
Then, we perform a period search procedure on the three time-series data: the ZTF-g band data, the ZTF-r band data, and the combined g + r band light curves.

Multiple algorithms can be used to the search of periodity from non-uniformly sampled data, including the \ac{LS} periodogram \footnote{\url{https://docs.astropy.org/en/stable/timeseries/lombscargle.html}} \citep{Lomb1976,Scargle1982,VanderPlas2018}, the \ac{CE} algorithm \citep{Graham2013a,Graham2013b,Katz2020}, and the \ac{BLS}  algorithm \citep{Kovcs2002,Shahaf2022}.

\textbf{(i) The Lomb-Scargle periodogram} is based on the Fourier transform and is used to detect and characterize the periodic component in unevenly sampled time-series data and generate a signal power spectrum \citep{VanderPlas2018}. The Lomb-Scargle normalized periodogram at frequency $f$ is defined as \citep{Lomb1976,Scargle1982,Leroy2012}.
\begin{equation}
      \label{eq:equation10}
\begin{aligned}
P_{N}(f)=& \frac{1}{2 \sigma^{2}}\left[\frac{\left[\sum_{k}\left(y_{k}-\hat{y}\right) \cos \omega\left(t_{k}-\tau\right)\right]^{2}}{\sum_{k} \cos ^{2} \omega\left(t_{k}-\tau\right)}\right.\\
&\left.+\frac{\left[\sum_{k}\left(y_{k}-\hat{y}\right) \sin \omega\left(t_{k}-\tau\right)\right]^{2}}{\sum_{k} \sin ^{2} \omega\left(t_{k}-\tau\right)}\right],
\end{aligned}
\end{equation}
where $\omega = 2\pi f$, $\sigma^{2}$ is the variance of the photometry, $y_{k}$ is the photometry at corresponding observation times $t_{k} (k = 1,...,N)$, $\hat{y}$ is the mean of the photometry, and $\tau$ is the time-offset which is defined by Eq. (3) from \citet{Leroy2012}.

\textbf{(ii) The Conditional Entropy algorithm} is a period-finding method to find the period of an astronomical (irregularly sampled) time series data based on minimizing the conditional Shannon entropy when the light curve folded at the trial period \citep{Graham2013a}. The expression of the conditional entropy $H_{c} (m_{i}\mid \phi_{j})$ can be defined as \citep{Graham2013a}.
\begin{equation}
      \label{eq:equation11}
      H_{c}\left(m_{i}\mid \phi_{j}\right) = \sum_{i,j} p(m_{i},\phi_{j}) \ln\left ( \frac{p(\phi_{j})}{p(m_{i}, \phi_{j})} \right),
\end{equation}
where $m$ is the normalized magnitude, $\phi$ is the phase (trial period), $p(m_{i},\phi_{j})$ is the density of points that fall within the bin located at phase $\phi_{j}$ and magnitude $m_{i}$, and p($\phi_{j}$) is the density of points that fall within the phi range.
The \ac{CE} algorithm determines the period through folding the phase of the light curve at each trial frequency, then estimating the conditional entropy of the partitioned phase-folded light curve corresponding to the trial frequency. In this paper, we used 10-magnitude bins and 20-phase bins in calculating the period of ZTF sources using the \ac{CE} algorithm.

\textbf{(iii) The Box-Least Squares algorithm} is based on a simplified box-shaped model of a strictly periodic transit, which is characterized by using only five parameter estimators to find the best model \citep{Kovcs2002}. The parameters used are the period ($P$), the transit duration as a fraction of the period ($q$),  the phase offset of the transit ($\phi_{0}$), the difference between the out-of-transit brightness and the brightness during transit ($\delta$), and the out-of-transit brightness ($y_{0}$).
The frequency spectrum of BLS can be defined by the amount of Signal Residue (SR) of the time series at any given trial period \citep{Kovcs2002}:
\begin{equation}
      \label{eq:equation12}
       {\rm SR} = \max \left \{ \left [ \frac{s^{2} }{r(1-r)}\right ]^{1/2} \right \},    
\end{equation}
where the equations of s and r are shown in \citet{Kovcs2002}, which can be used to estimate the photometric magnitude and the transit depth \citep{Panahi2021}. 
The light curve of eclipsing binaries may generate the secondary eclipse, however, the secondary eclipse is in some cases too shallow to be detected.
The \ac{BLS} algorithm searches for the periodic dips in brightness for extrasolar planets, especially for the secondary eclipse. 

The \ac{LS} algorithm has the advantage of being very fast to execute, but the \ac{CE} and the \ac{BLS} algorithms can report a more accurate period and are more robust against noises.
Therefore, we apply \ac{LS} period search to all 10938 sources, but only apply \ac{CE} and \ac{BLS} analysis if the sources show reliable periodic features. The search strategy of the period is separated into two stages: the rough search and the fine search.

\textbf{(1)} In the first stage of the rough search for the period, the LS algorithm can quickly calculate each source with have good ZTF light curve data in the Variable sample, however, LS has a different sensitivity to objects with eclipsing binaries and sinusoidal sources, and LS is more sensitive to the latter, which results in that eclipsing binaries have twice the real period, and there exists a peak at half of the true frequency in the frequency spectra, as shown in Figure \ref{fig:FSexample}.

\textbf{(2)} In the second stage of the fine search for the period, we visually inspected the phase-folded lightcurves of 10938 sources and determined that 826 of them are promising periodic sources (see Table \ref{tab:sampletype}), which we define as the \emph{periodic sample}.
In this search, we adopt the \texttt{cuvarbase} implementation of the \ac{BLS} and \ac{CE} algorithms, please see Appendix \ref{ref:appendixc} \footnote{\url{https://github.com/johnh2o2/cuvarbase}}.
A maximum frequency limit of 480 times per day is adopted.
For the minimum frequency, we adopt the limit as twice the inverse of the baseline, where the baseline is defined as the end time minus the start time of the time-series data (for further details, please see Appendix \ref{ref:appendixb} Period Finding \citet{Burdge2020a}).

\begin{figure*}[htpb]
\begin{center}
\includegraphics[width=1.0\textwidth]{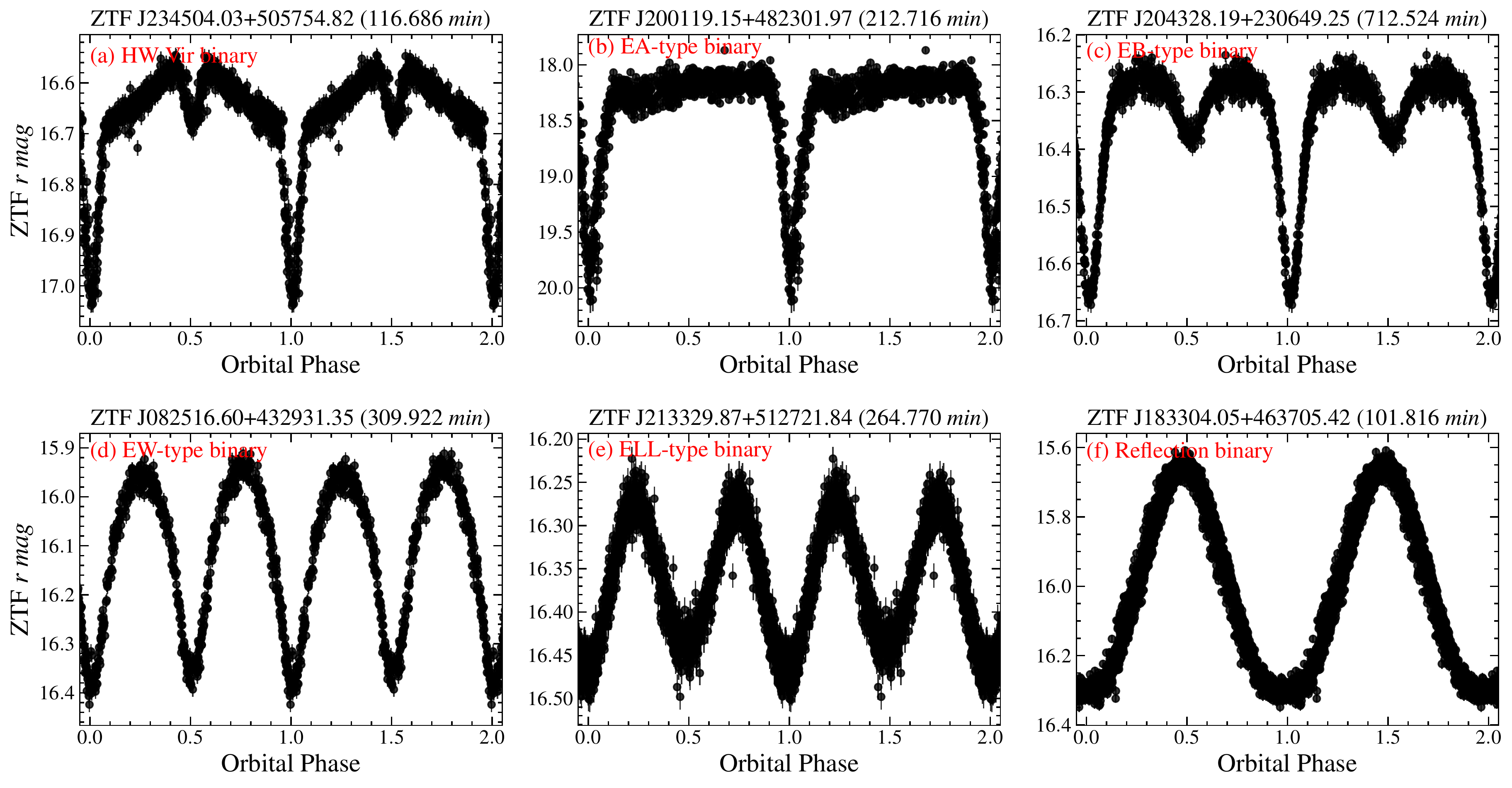}
  \caption{Examples of the variability types of light curve shapes caused by the binary eclipse effect in our binary sample. (a) HW Vir-type binary, in which a low-mass main-sequence star (M dwarf) orbits a core helium-burning sdO/sdB primary nearly edge-on \citep{El-Badry2021b}. (b) EA-type eclipsing binary, the unique shape of light curves is that between eclipses the light remains almost constant or varies insignificantly, and the light curves showing two minima per cycle or secondary minima may be absent. (c) EB-type eclipsing binary, showing two minima per cycle, and secondary minimum depth usually being considerably smaller than that of the primary minimum. (d) EW-type eclipsing binary, consisting of ellipsoidal components almost in contact, and the depths of the primary and secondary minima of the light curves are almost equal or differ insignificantly. (e) ELL-type binary, the light curve shape has the characteristics of quasi-sinusoidal variability produced by tidal deformation and shows two different minima in one orbit. (f) Reflection binary, in which an M dwarf orbits a hot white dwarf and is heated on one side \citep{El-Badry2021b}.}
\label{fig:LCexample}
\end{center}
\end{figure*}

\subsection{Periodic Sample}

The classification criteria based on the Variable Stars Index catalog\footnote{\url{https://www.aavso.org/vsx/index.php?view=about.vartypes}} can be divided into two main variability groups: extrinsic variable stars and intrinsic variable stars. Extrinsic variable stars can be divided into eclipsing, rotating (spots, reflections, and ellipsoidal shapes), and microlensing events, of which the first two groups have obvious orbital modulation characteristics.
The two main types of intrinsic variable stars are pulsation and outburst, in which pulsation is mainly the physical change inside the star.
Based on the classification criteria of variable stars, we studied the general characteristics of the light curve shapes for binary systems and expected the physical mechanism of the variations.

\begin{figure}
  \centering
  \includegraphics[width=\columnwidth]{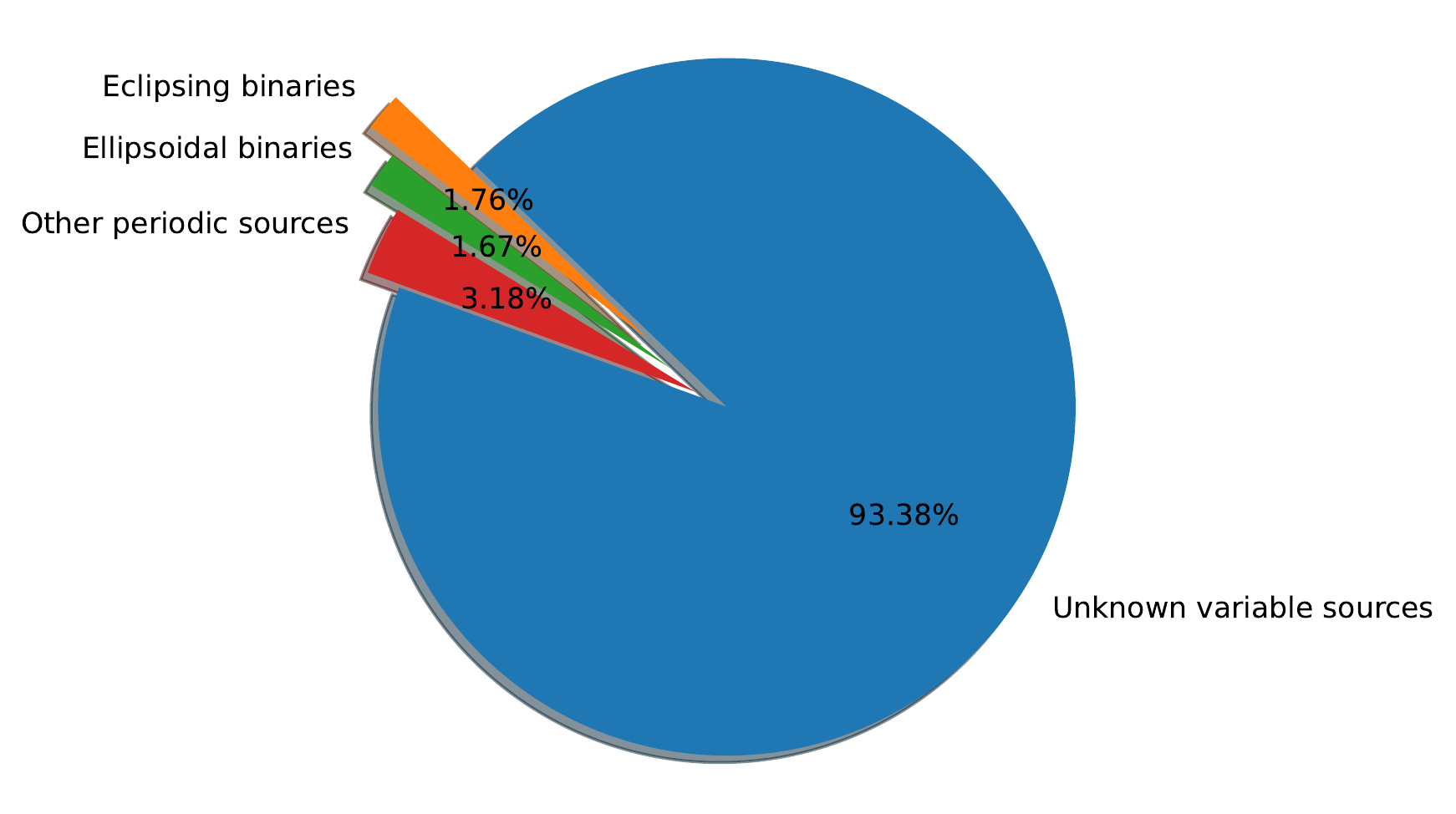}
  \caption{The pie chart of the percentage of binaries for particular variability classes in our variable sample.}
  \label{fig:PieVariable}
\end{figure}

Through multiple stages of search and veto, we are now left with a periodic sample of 826 candidates.
More than $93.38\%$ of the variable sample are discarded, as they demonstrate irregular or non-variability and outburst characteristics, as shown in Figure \ref{fig:PieVariable}.
The physical origin of outbursting stars could be similar to the dwarf nova (\ac{CV}) systems, especially the AM CVn-type systems \citep{Roestel2021,Marcano2021}.
However, we have no reliable spectroscopic follow-up observations to confirm the binary nature of the systems.
Therefore, we have to discard these sources for this study.

\begin{figure*}[htpb]
\begin{center}
\includegraphics[width=1.0\textwidth]{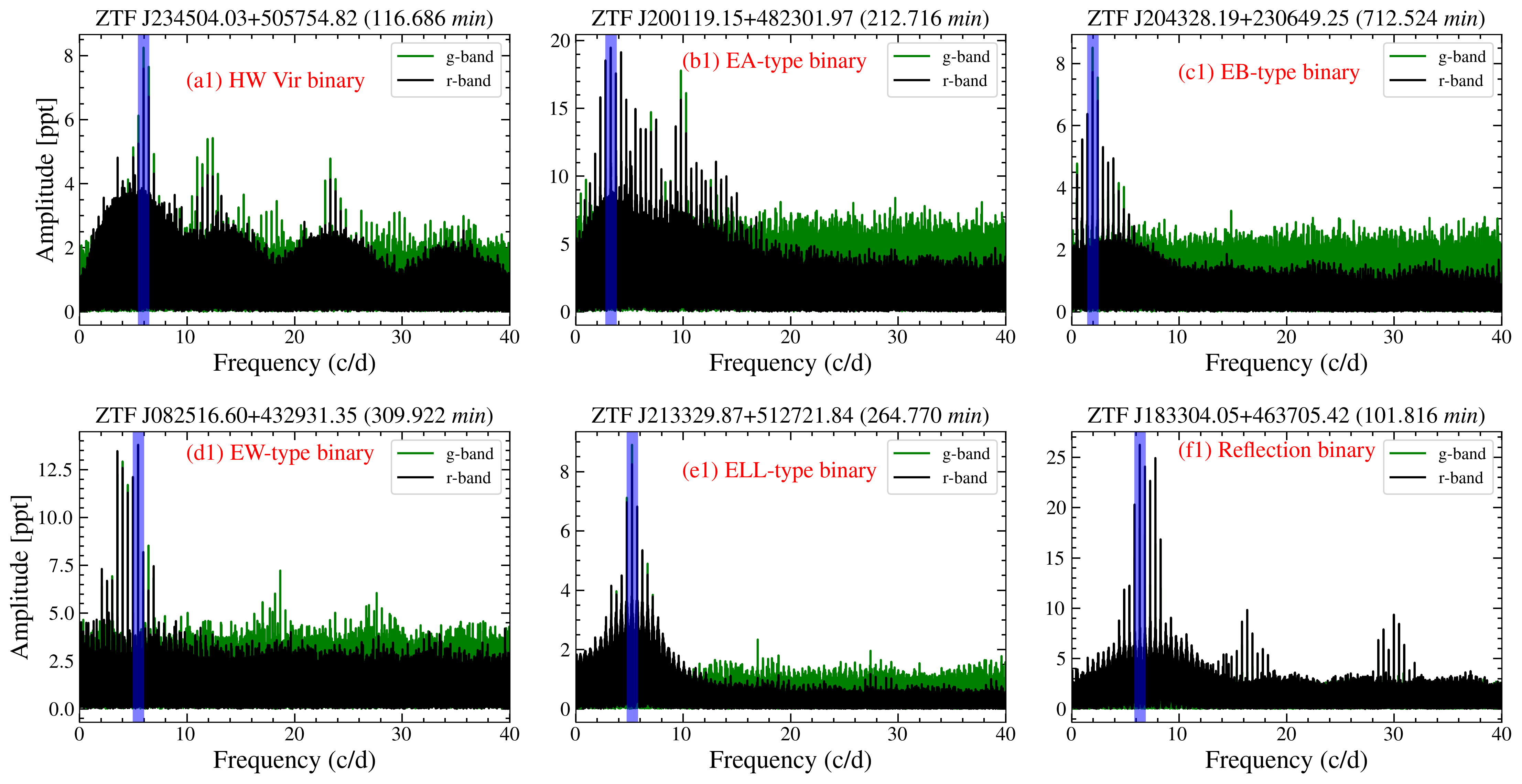}
  \caption{Examples of the variability types of light curve shapes caused by the binary eclipse effect in our binary sample. Panels (a1) to (f1) show the frequency spectra of the LS periodogram corresponding to the variability types of the phase folded ZTF light curve listed in Figure~\ref{fig:LCexample}.}
\label{fig:FSexample}
\end{center}
\end{figure*}

\section{Analysis and Results} \label{sec:analaysisresults}

\subsection{Binary Sample}

According to the light curve shapes, the 826 candidates from the periodic sample can be further classified as eclipsing binary systems, such as the HW Vir-type eclipsing binary, the EA-type eclipsing binary (Detached Algol-type binaries), the EB-type eclipsing binary ($\beta$ Lyrae-type binaries) and EW-type eclipsing binary (W Ursae Majoris type binaries), ellipsoidal binary system (ELL-type lightcurve binaries), reflection system (Figure~\ref{fig:LCexample} (f) show the typical light curve shape of reflection binary: ZTF J183304.05+463705.42.), pulsation (the RR Lyrae type pulsating star and the delta Scuti type pulsating star), and rotation sinusoids\footnote{\url{https://www.zooniverse.org/projects/ajnorton/superwasp-variable-stars/classify}}, etc.

The final selection is to pick up sources demonstrating the obvious binary feature, like orbital modulation characteristics by binary motion, or showing light curves consistent with ellipsoidal or eclipsing models (see Table \ref{tab:sampletype}).
Eventually, we produce a binary sample with 429 sources in total, as shown in Figure~\ref{fig:GaiaSampleHR} (d).
Among these, 220 are eclipsing binaries, and 209 are ellipsoidal binaries.

\subsubsection{HW Vir-type Binaries}

HW Virginis (HW Vir) systems are PCEBs composed of a sdO/sdB primary and a low-mass main-sequence star secondary (M-dwarf), e.g.~sdB+dM systems for the prototype. HW Vir-type binaries mostly show variations in their orbital periods ($<$0.1 days), also called eclipse time variations \citep[ETVs,][]{Sale2020}. 
The light curve of HW Vir systems shows that two distinct eclipses belong to the typical Algol-type eclipsing, but its light curve also has an out-of-eclipse variation, which is caused by the irradiation effect.
Figure~\ref{fig:LCexample} (a) show the typical light curve shape of HW Vir-type binary: ZTF J234504.03+505754.82 (116.686 min).
Among our eclipsing binary candidates, there are 58 HW Vir systems, accounting for 13.52$\%$ of all binary samples, as shown in Figure \ref{fig:PieBinary}. We present the result of Gaia EDR3 data information of these candidates in Table~\ref{tab:HWVirsources}, please see Appendix B.

\subsubsection{EA-type Binaries}

EA-type eclipsing binaries are also called detached Algol-type binaries. Based on the definition of the Variable Stars Index Catalog, the unique shape of light curves are that between eclipses the light curve remains almost constant or varies insignificantly. 
The explanation for this characteristic is that binaries have spherical or slightly ellipsoidal components, and the reflection effects or physical variations are also important factors.

Detached white dwarf binaries, can have different effective temperatures and can show eclipse variability in the photometric data.
If the binary has a high inclination angle, then the phase-folded light curve will contain two eclipses, a deeper primary one and a shallower secondary one.
The primary eclipse is generated when the brighter target is obscured by its companion, and the secondary eclipse is the opposite process.
In some cases, however, the secondary eclipse is too shallow to be detected, and there is only a primary eclipse in one orbit.
Out of 220 eclipsing binary candidates in the binary sample, 167 sources illustrate both primary and secondary eclipses, and only primary eclipses are identified for the remaining 59 sources.
Figure~\ref{fig:LCexample} (b) show the typical light curve shape of EA-type eclipsing binary: ZTF J200119.16+482301.88 (212.716 min).
Among our eclipsing binary candidates, there are 68 EA-type eclipsing binaries, accounting for 15.15$\%$ of all binary samples, as shown in Figure \ref{fig:PieBinary}. We present the result of Gaia EDR3 data information of these candidates in Table~\ref{tab:EAsources} in Appendix B.

\subsubsection{EB-type Binaries}

EB-type ($\beta$ Lyrae-type) eclipsing binary systems have ellipsoidal components and light curves for which it is impossible to specify the exact times of onset and end of eclipses because of a continuous change of the system's apparent combined brightness between eclipses. 
The secondary minimum is observed in all cases, its depth usually being considerably smaller than that of the primary minimum.
Figure~\ref{fig:LCexample} (c) show the typical light curve shape of EB-type eclipsing binary: ZTF J204328.19+230649.25 (712.524 min).
Among our eclipsing binary candidates, there are 56 EB-type eclipsing binaries, accounting for 13.05$\%$ of all binary samples, as shown in Figure \ref{fig:PieBinary}. We present the result of Gaia EDR3 data information of these candidates in Table~\ref{tab:EBsources} in Appendix B.

\begin{deluxetable*}{lcccc}[htpb]
\tablecaption{Summary of the numbers of close white dwarf binary candidates at different stages and different types. \label{tab:sampletype}}
\tablenum{1}
\tablewidth{0pt}
\tablehead{ \colhead{Sample} & \colhead{Type} & \colhead{Identified Sources}  & \colhead{Unidentified Sources}  & \colhead{Number} }
\startdata
$\mathbf{Selection~on~Gaia~data}$    &                                    &                    &                            &    \\       
$\emph{Initial sample}$                     &               $...$               &       $...$       &          $...$            & 823231 \\
$\emph{Variable sample}$                 &               $...$               &       $...$       &         $...$             & 12480  \\
$\mathbf{Selection~on~ZTF~data}$    &                                    &                    &                            &    \\       
$\emph{Periodic sample}$                &                $...$              &        $...$      &           $...$           &  826 \\
\hline 
\hline 
                                       & HW Vir-type (Algol-type) binaries                           & 6           & 52                  & 58 \\
                                       & EA-type (Detached Algol-type) binaries                  & 14          & 51                  & 65  \\
$\emph{Binary sample}$     & EB-type ($\beta$ Lyrae-type) binaries                   & 3           & 59                  & 56  \\
                                       & EW-type (W Ursae Majoris-type) binaries               & 1            & 40                  & 41  \\   
                                       & ELL-type (Ellipsoidal) binaries                               & 20          & 183                 & 209  \\
\hline 
          &           $\mathbf{Subtotal}$                                                &  $\mathbf{44}$         & $\mathbf{385}$                & $\mathbf{429}$ \\
\hline      
\enddata
\tablecomments{It is worth noting that we have statistics on the sources that have been confirmed by spectroscopic follow-up observations or spectral energy distribution analysis, which as Identified Sources.}
\end{deluxetable*}

\subsubsection{EW-type Binaries}\label{sec:EW_binary}

\begin{figure}
  \centering
  \includegraphics[width=\columnwidth]{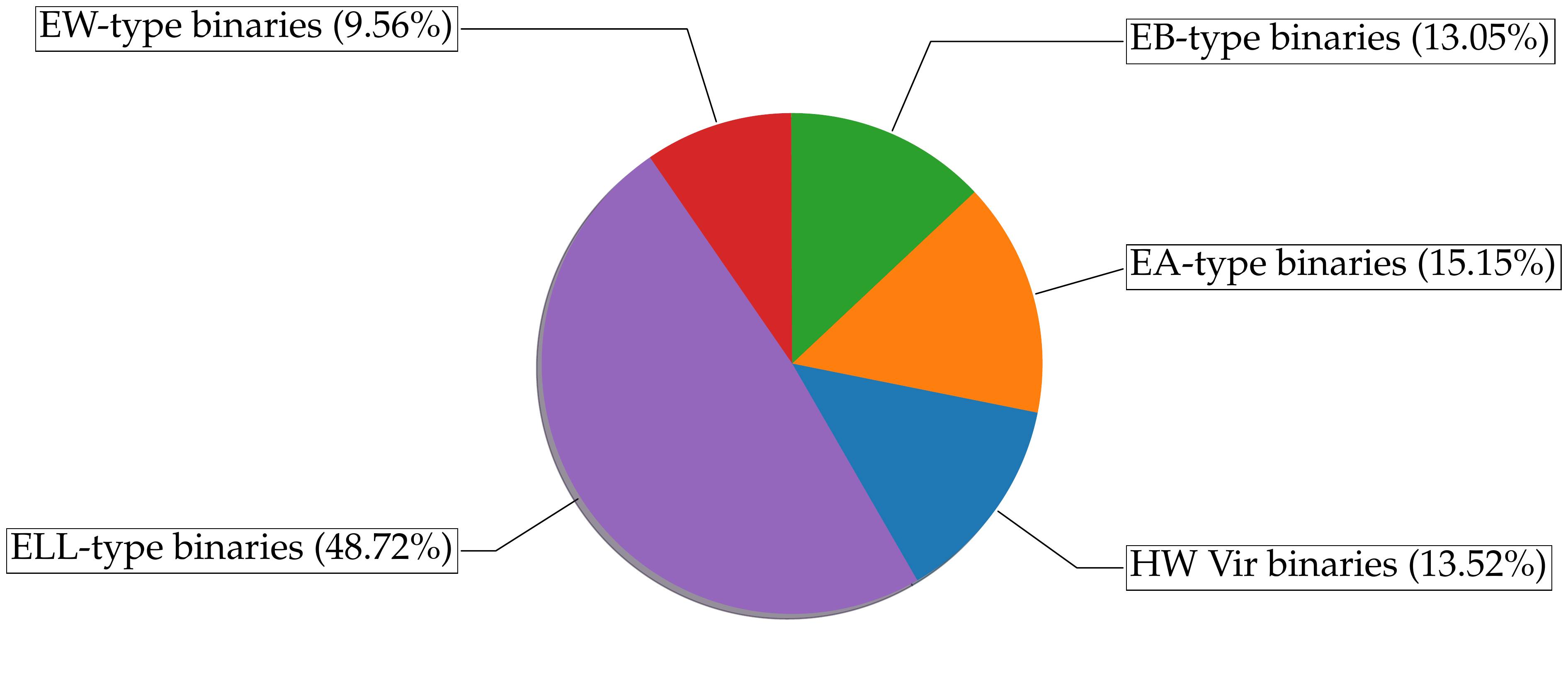}
  \caption{The pie chart of the percentage of binaries with the different types of light curve shapes in our binary sample.}
  \label{fig:PieBinary}
\end{figure}

EW-type eclipsing binaries are also known as W Ursae Majoris type binaries, which are composed of ellipsoidal components almost in contact, and eclipses in binary systems with components filling their Roche-lobes. 
The depths of the primary and secondary minima of the light curves are almost equal or differ insignificantly. 
The amplitude variations of light are usually less than 0.8 mag and the orbital periods are usually shorter than one day.
Figure~\ref{fig:LCexample} (d) show the typical light curve shape of EW-type eclipsing binary: ZTF J082516.60+432931.35 (309.922 min).
Among our EW-type eclipsing binary candidates, there are 41 EW-type eclipsing binaries, accounting for 9.56$\%$ of all binary samples, as shown in Figure \ref{fig:PieBinary}. We present the result of Gaia EDR3 data information of these candidates in Table~\ref{tab:EWsources}, please see Appendix B.

\subsubsection{ELL-type Binaries}

ELL-type binaries (ellipsoidal variables) are non-eclipsing reflection effect binaries consisting of a hot white dwarf and a cooler companion (typically an M dwarf) \citep{El-Badry2021b}. The variability amplitudes are usually less than 100 mmag.
In the ellipsoidal modulation system, the shape of the light curve has the characteristics of quasi-sinusoidal variability and shows two different minima in one orbit.
This quasi-sinusoidal variation light curve is produced by tidal deformation, in which case the two sides of the tidally disturbed star have different temperatures, and the hotter side is brighter than the colder side.
There are two different minima in the light curve, which is mainly caused by the gravity darkening effect.
Figure~\ref{fig:LCexample} (e) show the typical light curve shape of ELL-type eclipsing binary: ZTF J213329.86+512721.88 (264.770 min). 
We identify 209 candidates to be ellipsoidal variables, which account for 48.72$\%$ of the binary sample, as shown in Figure \ref{fig:PieBinary}.
We present the result of Gaia EDR3 data information of these candidates in Table~\ref{tab:ELLsources} in Appendix B.

\subsection{Comparison to the other catalogs}

Our screening scheme provides a candidate catalog covering different types of close white dwarf binaries, which provides observational evidence for the study of the evolutionary channels of white dwarf binaries.

Several studies feature the search of \ac{CWDB} systems.
We compare our samples and cross-check the identified binaries with these works.

\citet{Burdge2020b,Burdge2020a} used the Pan-STARRS1 catalog and ZTF photometric data to search for potential sources for milli-Hz band gravitational wave detectors.
They successfully discovered sixteen ultra-compact binaries, including eight eclipsing systems, two AM CVn systems, and six ellipsoidal variations systems.
Two of the sources, namely ZTF J213056.69+442046.58 (orbital period 39.340 min) and ZTF J205515.96+465106.45 (orbital period 56.348 min), are also identified in our search.
Both sources are mass-transferring WD+sdB systems, and their light curves have similar shapes, which are typical ellipsoidal variables (see \citet{Kupfer2019, Kupfer2020a, Kupfer2020b, Kupfer2022} for further details).
We analyzed the reason why only two sources in our sample overlap with the 15 ultra-compact binaries reported by \citet{Burdge2020b}. We checked the Gaia astrometric parameters of these 15 ultra-compact binaries and found that only three sources are measured with \texttt{parallax\_over\_error} $>$ 5, which means that most sources are filtered out by our initial selection criteria for lack of reliable distance estimation, see Section \ref{subsec:colorselection} for further details.

\citet{El-Badry2021b} selected sources in the Gaia color-magnitude diagram and searched for large-amplitude ellipsoidal variability using ZTF photometric data.
Their motivation is to find the progenitor of extremely low-mass white dwarfs and AM CVn systems.
In \citet{El-Badry2021b}, the authors reported 51 candidates, of which 21 sources obtained many-epoch spectra, and all 21 sources were confirmed to be completely or nearly Roche lobe filling binaries, 13 showing evidence of ongoing mass transfer.
Our 22 ellipsoidal variable candidates reported by \citet{El-Badry2021b}, and 11 sources have been confirmed by spectra.
The initial query conditions we used for sample selection in Gaia EDR3 are different from \citet{El-Badry2021b}.

\citet{Keller2022} used the published catalog compiled by \citet{Fusillo2019} to identify eclipsing white dwarf binaries.
They used the \ac{BLS} algorithm to search for sources with periodic light curve variability in the ZTF data, and their search revealed 18 new binaries.
The cross-check between our binary sample and these 18 binaries reveals five binaries to be overlapped.
\citet{Keller2022} used samples from the Gaia white dwarf catalog, which contains 486641 sources. 
They cross-matched Gaia white dwarf catalog with ZTF DR3 data and finally obtained 276074 sources that have at least two ZTF epochs.
We used the {\emph{variable sample}} which was selected after the Gaia Variability Metric cut to cross-match the ZTF DR8 data.
This leads to the discard of a large fraction of faint sources and could explain the difference in reported samples.

\citet{WangKevin2021} systematically searched for periodic variables in the hot subdwarf catalog from Gaia DR2 using ZTF data.
Their targets come from a catalog of 39,000 hot subdwarf candidates provided by \citet{Geier2019}.
In their search, they found 67 HW Vir binaries and 496 sources with reflection effects or pulsation, plus a few eclipsing and ellipsoidally modulating binaries.
Our selection criteria in Gaia color-magnitude diagram include the hot subdwarf region in Section 3.2 of \citet{Geier2019}.
Therefore, our candidates will inevitably overlap with \citet{WangKevin2021}.
However, the catalog is not publicly available, therefore we can not check the extent of overlap.

\begin{figure*}[htpb]
\begin{center}
\includegraphics[width=0.85\textwidth]{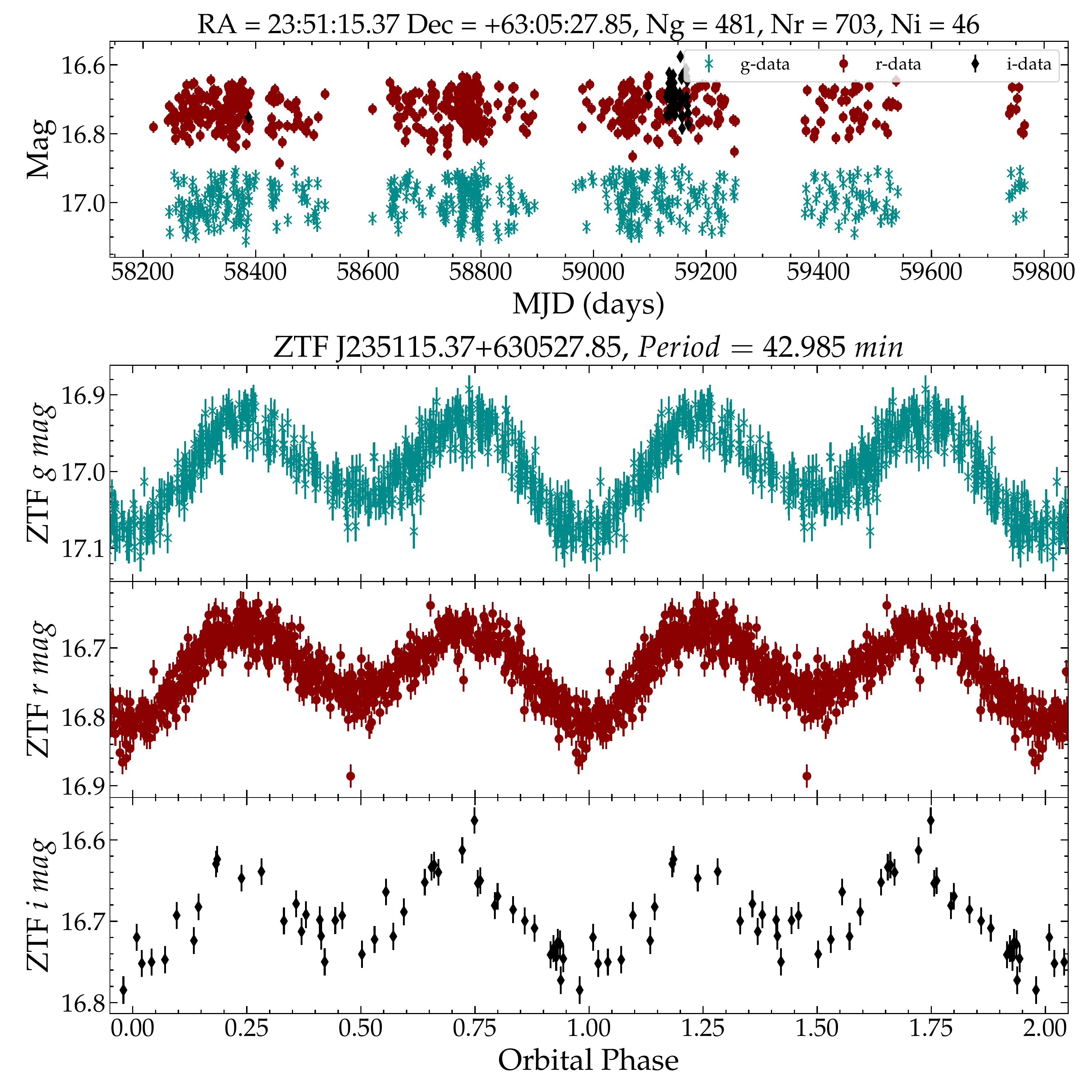}
  \caption{The light curves of ZTF J235115.32+630528.23. The top panel shows the raw ZTF g (dark cyan), r (dark red), and i (black) light curves. The middle and the bottom panel show the ZTF g-band and r-band folded light curves, respectively. Archival ZTF g-band and r-band light curves of the system folded at a period of 42.985 minutes.}
\label{fig:ZTFJ235115}
\end{center}
\end{figure*}  

\begin{figure*}[htpb]
\begin{center}
\includegraphics[width=0.85\textwidth]{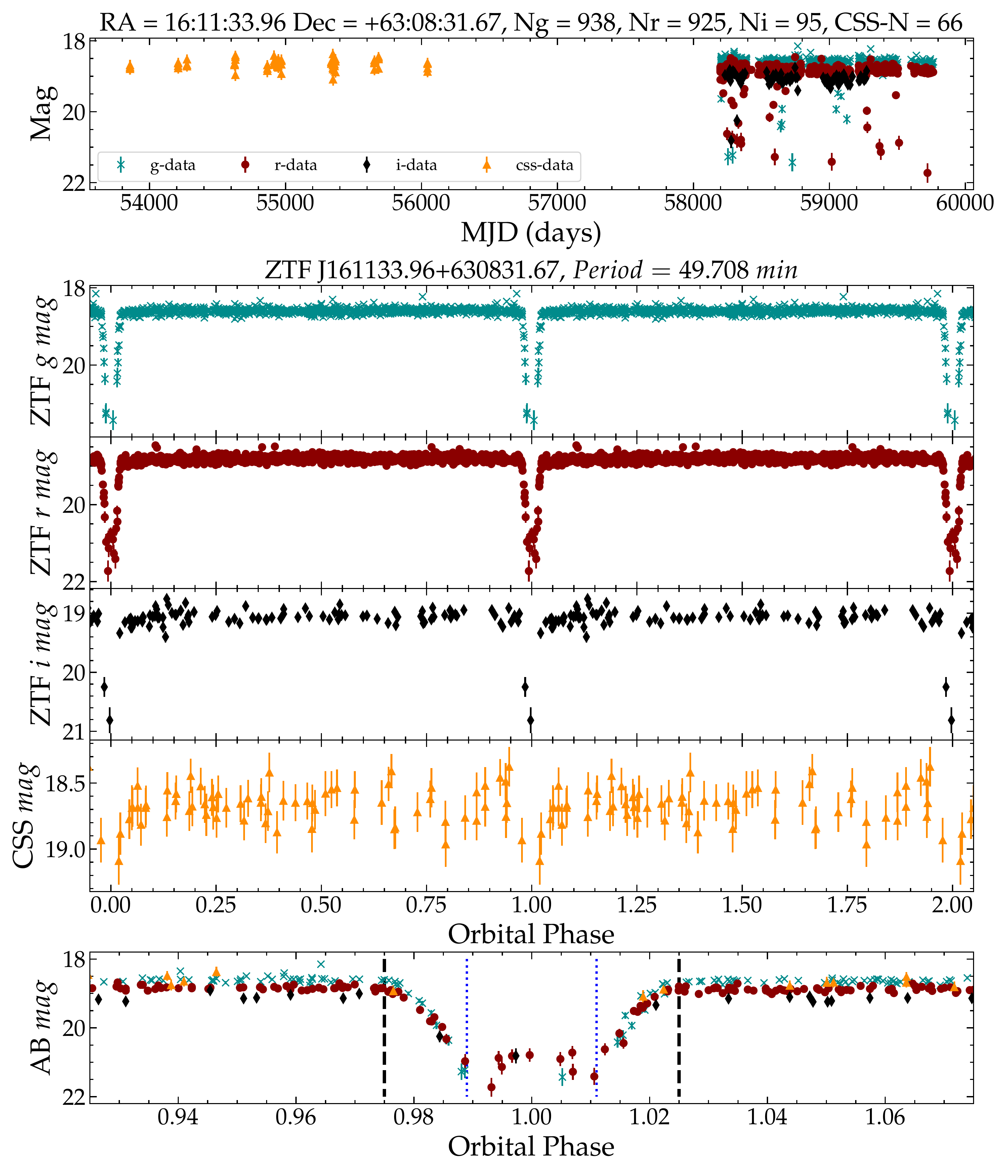}
  \caption{The light curves of ZTF J161133.96+630831.67. The top panel shows the raw ZTF g-band (dark cyan), r-band (dark red), and i-band (black) light curves, and raw CSS (dark orange) photometric data. The middle panels show the phased ZTF light curves for the g-band, r-band, and i-band, and the phased CSS light curve, respectively. The bottom panel shows the phase-folded light curves of ZTF J1611+6308 at phase $\phi \sim$ 0.975-1.025 in the eclipse duration. The blue dashed lines show the duration of the conjunction of the eclipse, and the black dashed lines show the entire duration of the eclipse from ingress to egress. Archival ZTF (g-, r-, and i-band), and CSS light curves of the system folded at a period of 49.708 minutes.}
\label{fig:ZTFJ161133}
\end{center}
\end{figure*}  

\begin{figure*}[htpb]
\begin{center}
\includegraphics[width=0.85\textwidth]{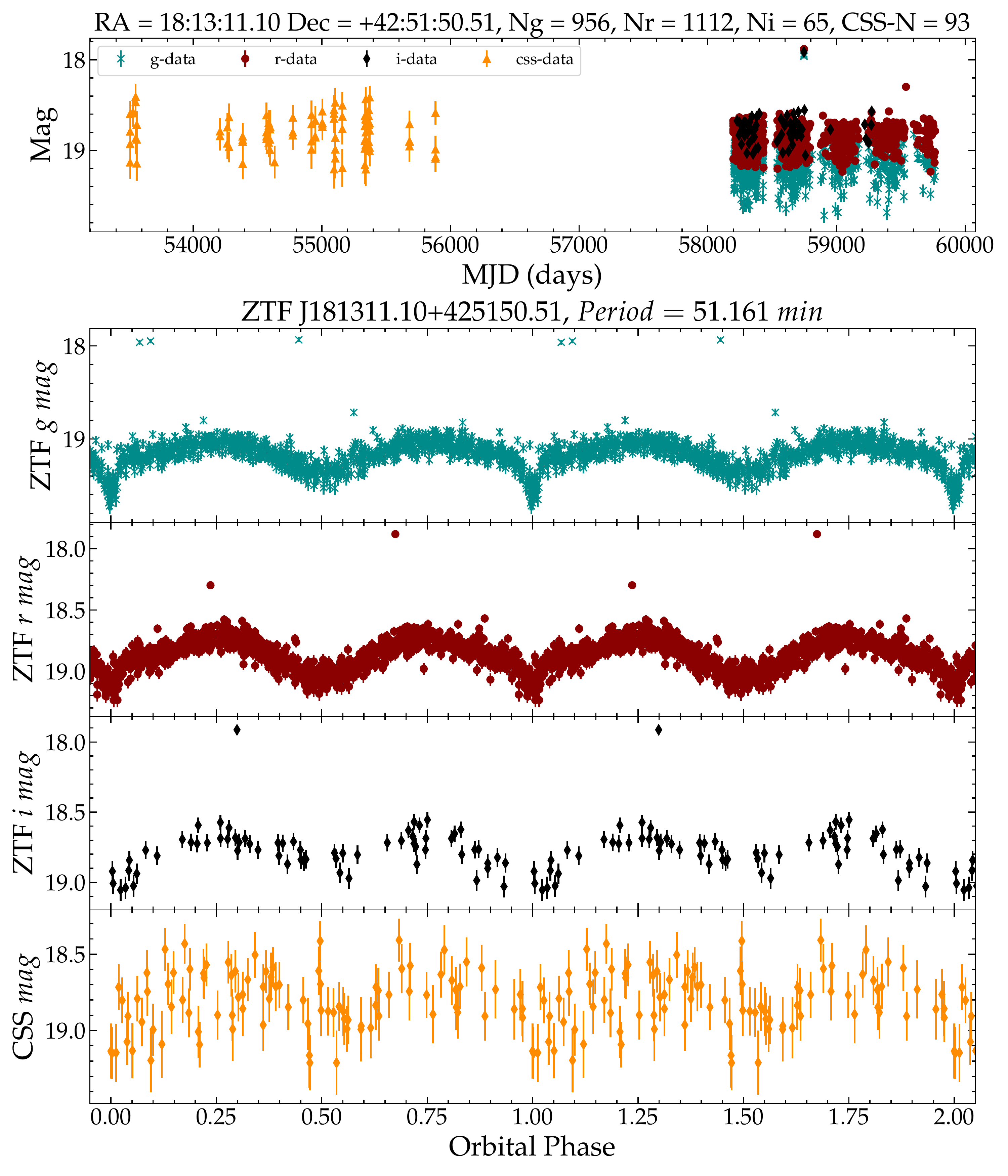}
\caption{The light curves of ZTF J181311.10+425150.51. The top panel shows the raw ZTF g-band (dark cyan), r-band (dark red), and i-band (black) light curves, and raw CSS (dark orange) light curves. The middle and the bottom panels show the phased ZTF (g-band, r-band, and i-band), and phased CSS light curves for ZTF J1813+4251, respectively. Archival ZTF (g, r, and i), and CSS light curves of the system folded at a period of 51.161 minutes.}
\label{fig:ZTFJ181311}
\end{center}
\end{figure*}  

\begin{figure*}[htpb]
\begin{center}
\includegraphics[width=0.85\textwidth]{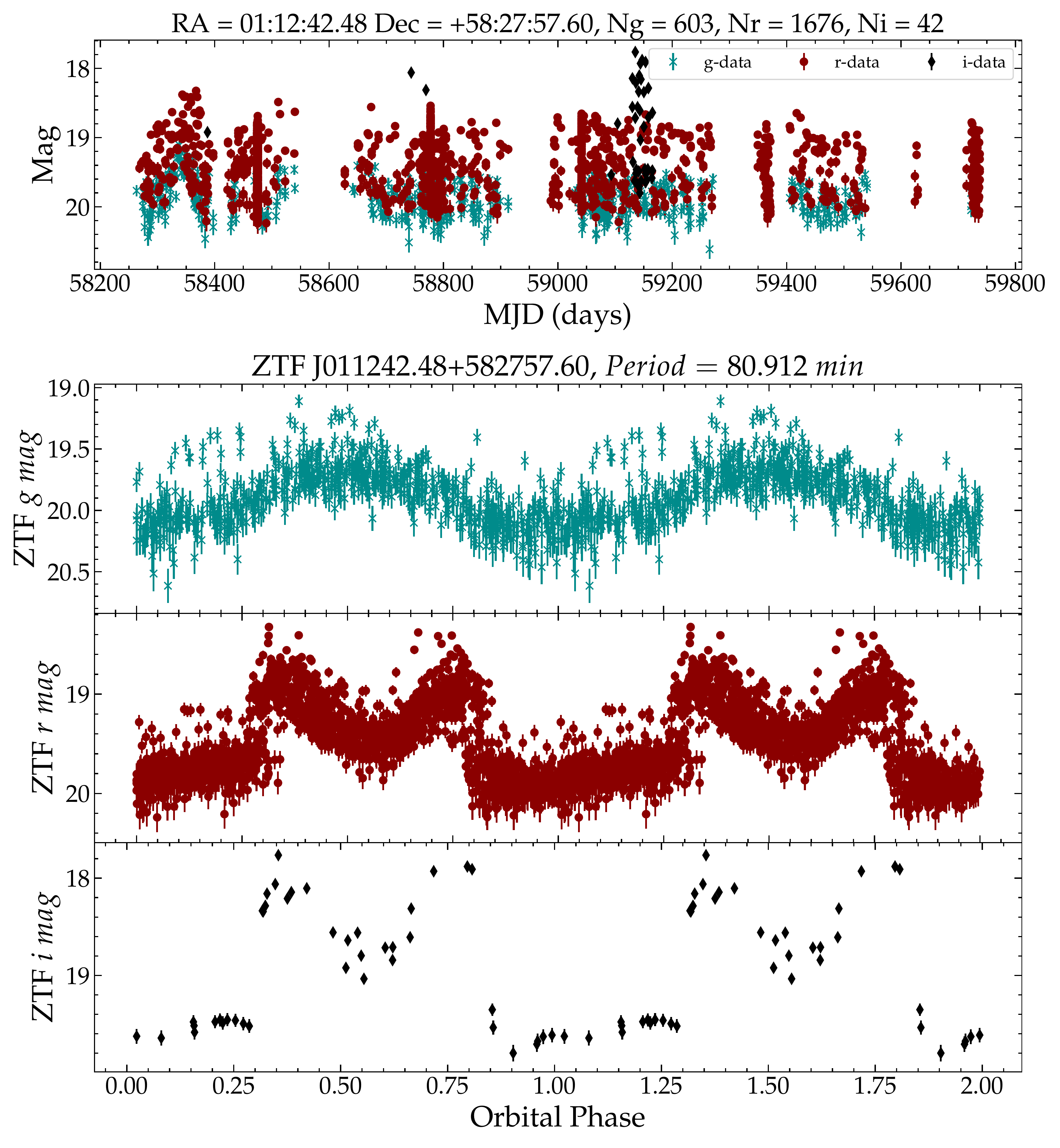}
\caption{The light curves of ZTF J011242.48+582757.60. The top panel shows the raw ZTF g-band (dark cyan) and r-band (dark red) light curves. The middle and bottom panels show the phase-folded light curves for ZTF J0112+5827 in g and r-bands, which show a clear difference between the two bands. The r-band exhibit two obvious spikes in one phase, between two spikes there is a transit phase.}
\label{fig:ZTFJ011242}
\end{center}
\end{figure*}

\section{Discussion} \label{sec:discussion}

\subsection{Notes on Individual Objects}
We selected four sources with special light curve shapes from the candidates, which are ZTF J2351+6305, ZTF J1611+6308, ZTF J1813+4251, and ZTF J0112+5827.
The four newly discovered \ac{CWDB} candidates have distinct light curve variability.
Three sources are accompanied by an orbital period of less than 60 minutes and that of the other source is about 81 minutes.
To understand the composition of these candidates and to explain the feature of the light curves, we will apply a large aperture telescope for time-domain photometry and spectral follow-up observations.

\textbf{ZTF J235115.32+630528.23} (ZTF J2351+6305) is an ultra-short period binary candidate.
Its phase-folded light curve shows ellipsoidal variation characteristics (as shown in Figure~\ref{fig:ZTFJ235115}), and the \ac{BLS} period is 42.985 minutes.
It has a color of $G_{\rm BP}-G_{\rm RP}=0.59$ mag and a magnitude of $G_{\rm abs}=6.24$ mag. 
It may be a mass-transferring and recently-detached CV, which is a progenitor of ELM-WDs or AM CVn systems.
\citet{El-Badry2021b} has systematically studied such systems.
Among their 51 candidates of evolved CVs and bloated proto-ELM WDs, the shortest orbital period is 2 hours.
This candidate has a shorter orbital period than reported by \citet{El-Badry2021b}.
The donor of evolved CVs has high temperatures ($4700 < T_{\rm eff} < 7000 \rm K$), which can be confirmed by spectral follow-up observation.

\textbf{ZTF J161133.96+630831.67} (ZTF J1611+6308) has a period of 49.708 min.
The ZTF phase-folded light curve shows a deeply-eclipsing white dwarf binary characteristic and the eclipse depth is greater than 2.5 mag, as shown in Figure~\ref{fig:ZTFJ161133}.
ZTF J1611+6308 has a color of $G_{\rm BP}-G_{\rm RP}=0.19$ mag and magnitude of $G_{\rm abs}=11.26$ mag.
The Catalina Sky Survey \footnote{\url{https://crts.iucaa.in/CRTS/}} observed ZTF J1611+6308 between 2005 and 2013, and we obtained clean raw photometric data of CSS for ZTF J1611+6308.
Figure~\ref{fig:ZTFJ161133} shows the phased ZTF light curves for the g-band, r-band, and i-band, and the phased CSS light curve.
The light curve has a deep eclipse, but there is no secondary eclipse, and the eclipse duration is about 2.5 minutes as shown in the bottom panel of Figure~\ref{fig:ZTFJ161133}.
We used twice the period to fold the light curve, and the primary and secondary eclipses appeared at the same depth.
The evolution of the light curve is predicted to be dominated by a white dwarf, while the companion star is a low luminosity target, such as a brown dwarf or an M dwarf \citep{Kosakowski2022}.
The classification of the system can be further confirmed by spectrum and high-speed photometric observation.

\begin{figure*}[htpb]
\begin{center}
\includegraphics[width=0.95\textwidth]{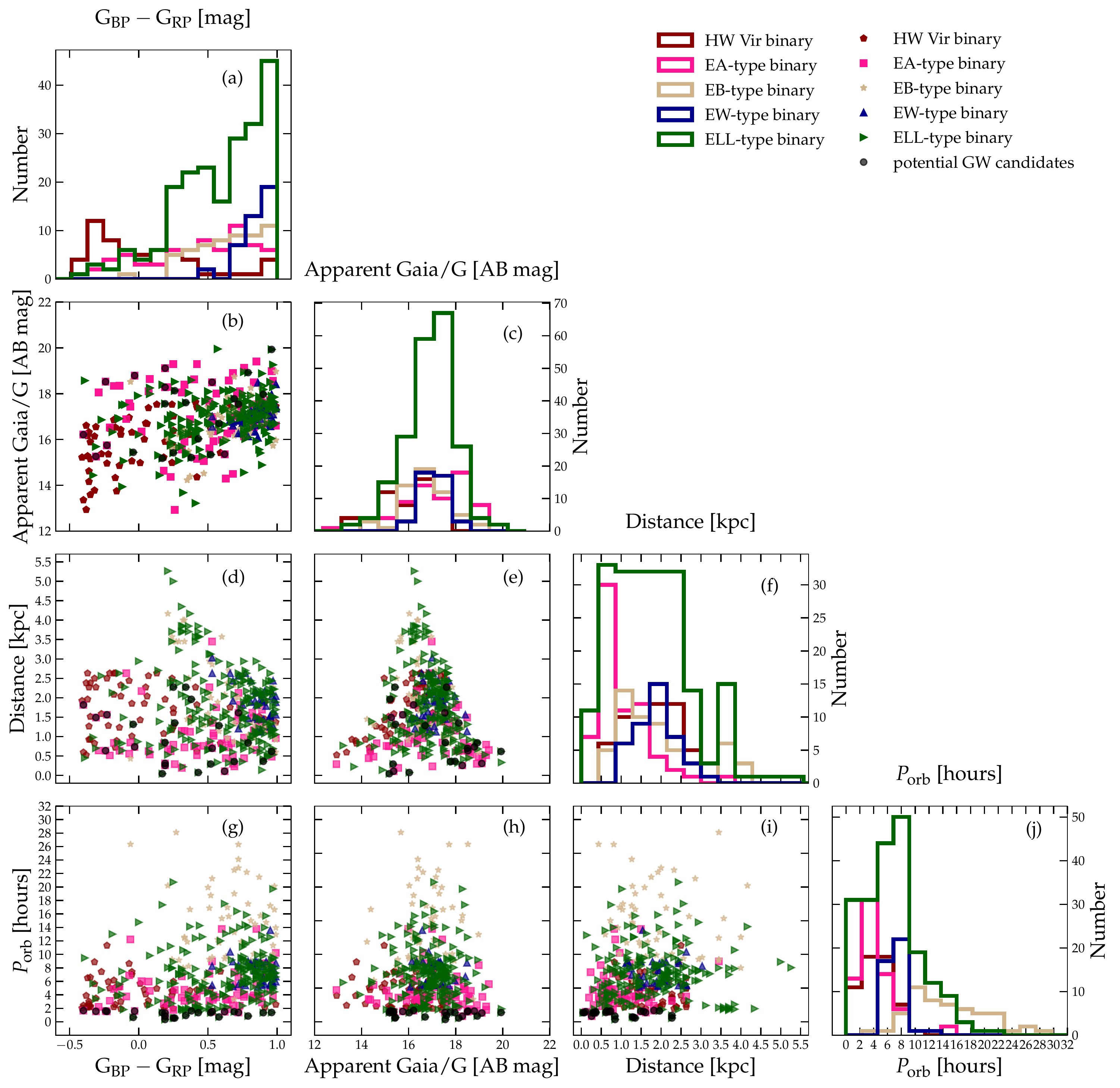}
\caption{The corner plot for the main parameters of different types of light curve shapes of binaries within our \emph{binary sample}. In panel (a), (c), (f), and (j), we show the one-dimensional histogram over Gaia \texttt{BP-RP} color, the Gaia G-band apparent magnitude, the distance, and the orbital period, respectively. Other panels (b, d, e, g, h, i) show the correlation distribution between the two parameters for different types of binaries.}
\label{fig:Distribution}
\end{center}
\end{figure*}  

\textbf{ZTF J181311.10+425150.51} (ZTF J1813+4251) is a cataclysmic variable candidate with a period of 51.161 min.

The light curve is a fully eclipsing binary system characteristic that is similar to the ZTF J1946+3203 \citep{Burdge2020a}, which exhibits strong ellipsoidal variations owing to the tidal deformation and eclipse effect as shown in Figure~\ref{fig:ZTFJ181311}.
ZTF J1946+3203 is a single-lined spectroscopic eclipsing binary, which is composed of a hot He WD or an sdB and a low luminosity He WD companion star.
In the color-magnitude diagram, ZTF J1813+4251 is located in the classical CV region at $G_{\rm BP}-G_{\rm RP}=0.79$ mag, and an absolute magnitude of $G_{\rm abs}=9.36$ mag, which is significantly different from the position of ZTF J1946+3203.
We obtained clean raw photometric data of CSS and ZTF for ZTF J1813+4251, as shown in the upper panel of Figure~\ref{fig:ZTFJ181311}.
Further observations of ZTF J1813+4251 are needed to measure the radial velocity and to strongly constrain the masses, radii, and temperatures using high-speed photometry.
Once confirmed, this may be the shortest orbital period CV system ever observed. 
During the preparation of this manuscript, \citet{Burdge2022} reported the observation of this source. They found that ZTF J1813+4251 is a transitional CV, and this binary consists of a star with a temperature comparable to that of the Sun but a density 100 times greater owing to its helium-rich composition, accreting onto a white dwarf. These transitional CVs have been proposed as progenitors of helium CVs \citep{Burdge2022}.

\begin{figure*}[htpb]
\begin{center}
\includegraphics[width=0.90\textwidth]{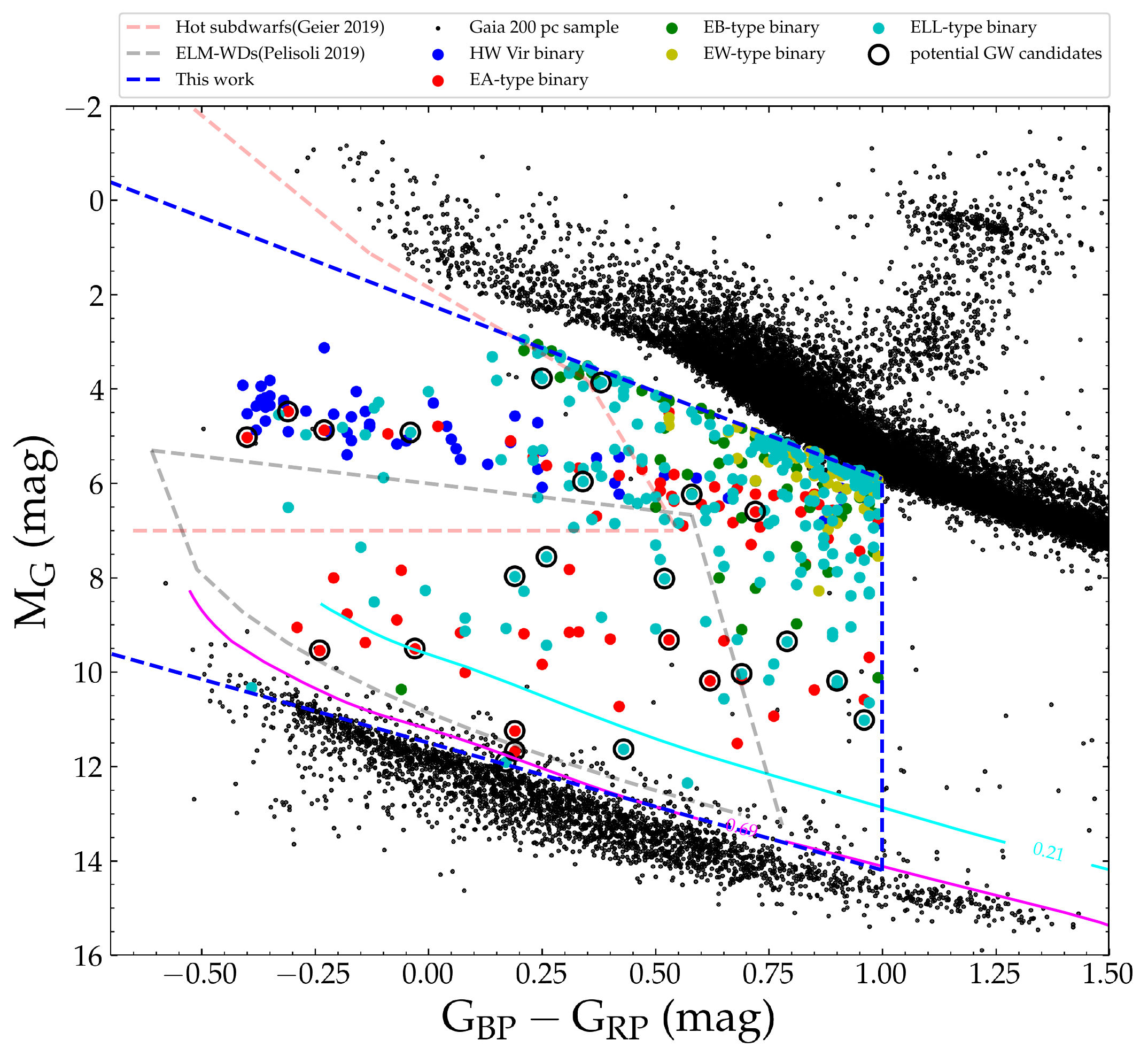}
\caption{The location of the binary samples on the Gaia H-R diagram. The blue dashed lines show our selection scheme for searching CWDBs based on Gaia EDR3 data. The red dashed lines show that the selection rule proposed by \citet{Geier2019} based on Gaia DR2 data to search the population of hot subdwarfs. The black dashed lines show that the selection scheme proposed by \citet{Pelisoli2019b} based on Gaia DR2 data to search ELM-WD candidates. The solid red line is the cooling model offset by $- 0.75$ mag for 0.21$M_{\odot}$ and 0.69$M_{\odot}$ white dwarf. We use the publicly available \texttt{WD\_models} package provided by \citet{Cheng2020}. The black scatter points show the Gaia 200 pc background sources. The five colors scatter points show the different types of light curve shapes in the final binary samples searched using ZTF data, and the black circles show the binaries in the \emph{Binary sample} whose period is less than 100 min, which are potential \ac{GW} source candidates.}
\label{fig:CWDGaiaHR}
\end{center}
\end{figure*}  

\begin{figure*}[htpb]
\begin{center}
\includegraphics[width=0.90\textwidth]{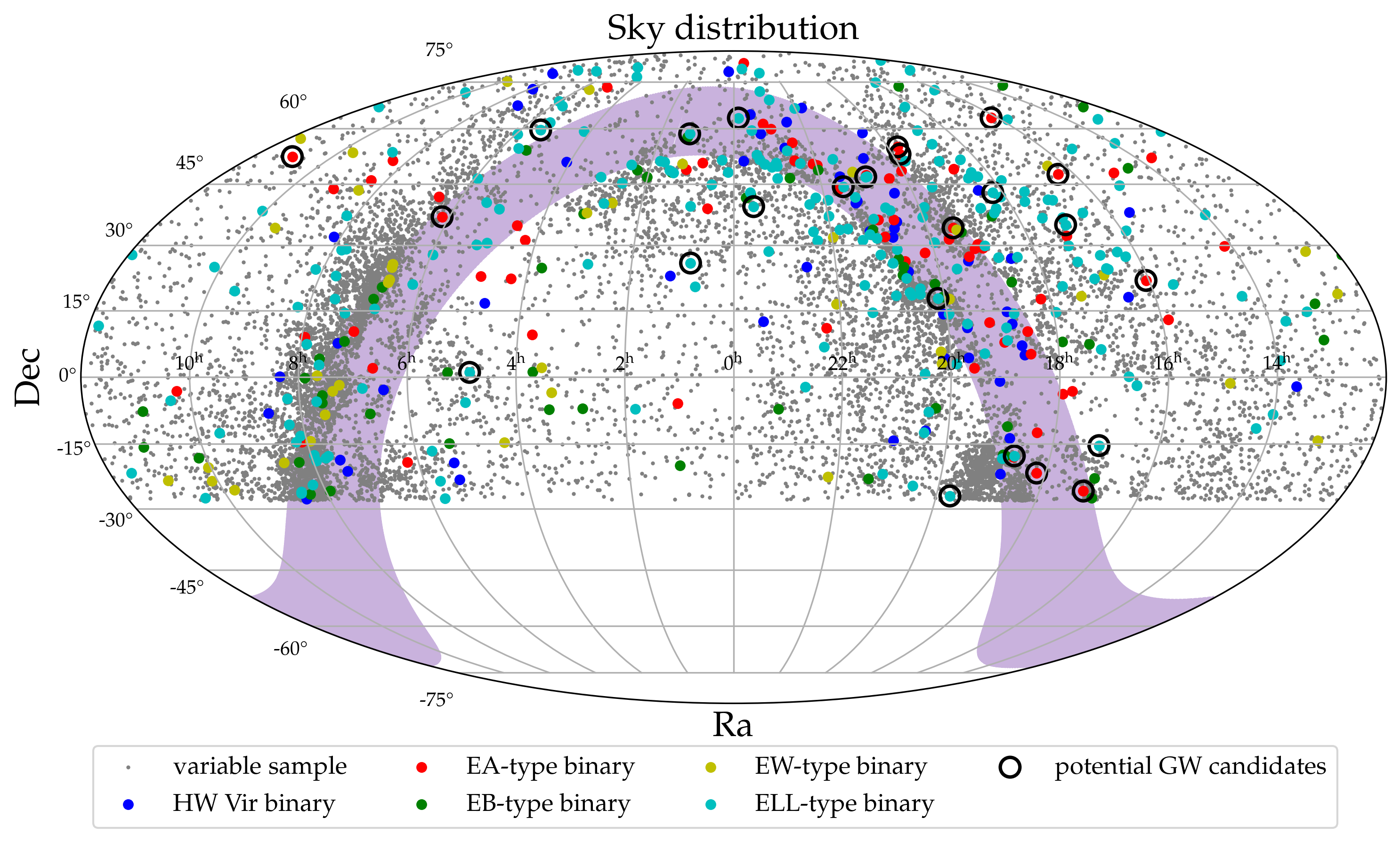}
\caption{Sky distribution in equatorial coordinates of the \emph{variable sample} (gray dots) and the \emph{Binary sample} (color dots). The black circles show the potential GW source candidates.}
\label{fig:Skymap}
\end{center}
\end{figure*}  

\textbf{ZTF J011242.48+582757.60} (ZTF J0112+5827) is a CV candidate, which has a color of $G_{\rm BP}-G_{\rm RP}=0.9$ mag and magnitude of $G_{\rm abs}=10.19$ mag.
The light curve of ZTF J0112+5827 shows typical large-amplitude ellipsoidal variation in the g-band, and the period computed by the BLS algorithm is 80.9 minutes, however, its r-band light curve shows an unusual two spikes in one period, as shown in Figure~\ref{fig:ZTFJ011242}.
We propose a possible hypothesis to explain the unusual light curve of ZTF J0112+5827:
This is a polar (AM Her-type star) or intermediate polar\footnote{\url{https://asd.gsfc.nasa.gov/Koji.Mukai/iphome/iphome.html}}.
The accreted mass moves along the magnetic lines in the direction of the white dwarf's magnetic poles.
The two spikes of the light curve in one period are generated by the accretion point, which is similar to a rare eclipsing polar BS Tri system \citep{Kolbin2022}.
Future high-time resolution spectral observations will be needed to verify the formation of the unusual light curve of the binary system.

\subsection{Parameters Distribution of CWDBs} \label{subsec:parameters}

In order to better understand the distribution characteristics of photometric parameters and source parameters of different types of candidates in our binary sample, we display their statistical histograms and correlations in Figure \ref{fig:Distribution}. 

Figure \ref{fig:Distribution} (a) shows the Gaia \texttt{BP-RP} color index distribution for different types of binaries. HW Vir binaries are mainly distributed at \texttt{BP-RP} $<$ 0 mag, because hot subdwarfs have a high surface temperature and they appear to be bluer. Other variability types of binaries are mainly distributed at \texttt{BP-RP} $>$ 0 mag, where ELL-type binaries are generally distributed between 0.2 mag $<$ \texttt{BP-RP} $<$ 1.0 mag. Figure \ref{fig:Distribution} (c) shows the apparent magnitude distribution of Gaia/G-band of different types of candidates, which scatters between 16-18 mag. In Figure \ref{fig:Distribution} (f), we show the distance distribution of all binary candidates, HW Vir binary and EW-type binaries are mainly between 1.0 kpc to 2.5 kpc, EA-type binary at 0.5-2.0 kpc, the EB-type binary at 1.0-2.0kpc. Figure \ref{fig:Distribution} (j) shows the orbital period distribution of these binary candidates. We find that for most types of binaries, the orbital periods are mostly less than 9 hours, while EB-type binary exhibits a peak at around 10 hours.  
We can explain the period distribution by the fact that the viewing angle will be decreasingly small for wider seperated eclipsing binaries. The ELL-type does not subject to this selection bias as they do not rely on eclipsing observation. The HW-type are also not affected as the component size are usually larger, corresponding to a larger viewing angle.

Other panels (b, d, e, g, h, i) in Figure \ref{fig:Distribution} show two dimensional distribution for different types of binary systems. For example, from Figure \ref{fig:Distribution} (d) and (e), we find an apparent cut in the upper right corner. This is most likely due to the choice of the Gaia variability metric.

\subsection{The Gold Sample of CWDBs}  \label{subsec:goldsample}

In the H-R diagram, the sub-types of \ac{CWDB} are mostly located in the region between MS and WDCS.
We divided the volume according to the types of close binary stars in the H-R diagram and defined three gold samples, namely PCEBs gold sample, CVs gold sample, and DWDs gold sample.
\citet{Inight2021} established a volume-limited sample for all sub-types of \ac{CWDB}, which can be used to study the population of WD binaries.
The reference samples provide four types of gold samples: WD + M (M-type main-sequence star), WD + AFGK (A, F, G, or K-type star), CV, and DWD.
Our gold samples are not established by spectral type.
We define the gold samples based on Gaia selection source conditions proposed by \citet{Geier2019} and \citet{Pelisoli2019b} and combined them with the white dwarf cooling model.
We show the distribution characteristics of these binary candidates in the H-R diagram in Figure \ref{fig:CWDGaiaHR}, and their sky distribution in Figure \ref{fig:Skymap}.

\citet{Geier2019} applied a color-cut and absolute magnitude selection to define a distribution range of hot subdwarf candidates in Gaia DR2, as shown in Figure~\ref{fig:CWDGaiaHR} (the red dashed lines). We further define the sdB binaries (or PCEBs) in this region as the gold sample (see Table \ref{tab:goldsample}). We have a total of 85 hot subdwarf binary candidates as gold samples, of which 6 have orbital periods less than 100 min. 

\begin{deluxetable}{ccc}[htpb]
\tablecaption{Summary of the differences in the number of the gold sample
\label{tab:goldsample}}
\tablenum{2}
\tablewidth{0pt}
\tablehead{ \colhead{Object Type} & \colhead{Total} & \colhead{$\mathrm P_{\rm orb} \leq$ 100~min candidate} }
\startdata
        PCEB Gold Sample & 85 & 6  \\
        CV Gold Sample & 224 & 7  \\
        ELM-WD Gold Sample & 45 & 9  \\
        DWD Gold Sample & 11 & 6  \\
\enddata
\tablecomments{The above four Gold samples are empirically based on the distribution of candidates in the Gaia H-R diagram. There have crossover regions in the segmentation criteria of the four Gold samples, which result in repeated statistics for some sources, for further details , see Figure~\ref{fig:CWDGaiaHR}.}
\end{deluxetable}

The CV gold sample established by \citet{Inight2021} shows that it is widely distributed below the MS star and the \texttt{BP-RP} colour index \texttt{$\rm G_{BP} - G_{RP}$} $>$ 0 in the Gaia H-R diagram (the detailed analysis is shown in Section 4.3 of \citet{Inight2021}). 
From the point of view of the time-domain photometry observation, it can be found that the light curve of \ac{CV} can be divided into three main types: ELL-type binary, EA-type binary, and the outbursting class. 
Our short-period \ac{CWDB} catalog only retains the EA-type or ELL-type binaries formed by the transit process, excluding the flare sources of the irregular class. Most CVs have orbital periods $P_{\rm orb} \lesssim$ 14 h. We have 224 candidates as CV gold samples in our catalog (see Table \ref{tab:goldsample}), of which 7 have periods less than 100 minutes.

\citet{Pelisoli2019b} proposed a method to select ELM candidates catalog from Gaia DR2 data, which is based on the distribution of the known samples in the Gaia H-R diagram and the prediction of the theoretical model. We define the \ac{CWDB} candidates in Pelisoli's color-cutting criterion as the gold samples of ELM binaries (see Table \ref{tab:goldsample}). In our \ac{CWDB} catalog, 45 candidates belong to the gold sample of ELM binaries, of which 9 have orbital periods less than 100 minutes.
A binary system consisting of two white dwarfs has a color similar to that of a single white dwarf, but a DWD system is brighter than a single WD. One can expect that DWDs appear above single WDs (up to $\simeq$0.75 mag) in the H-R diagram. In Figure~\ref{fig:CWDGaiaHR}, we use the cooling model contours ($- 0.75$ mag) of 0.2~$\mathrm M_{\odot}$ and 0.69~$\mathrm M_{\odot}$ single WDs to define the gold sample of DWDs. We use the publicly available \texttt{WD\_models} package provided by \citet{Cheng2020}. The \texttt{WD\_models} package is available at GitHub.\footnote{\url{https://github.com/SihaoCheng/WD_models}}. According to the definition of the DWD gold sample, we have 11 candidates, including 6 ultra-short periods ($P_{\rm orb} \leq$ 100~min) DWDs.

\begin{figure*}[htpb]
\begin{center}
\includegraphics[width=0.90\textwidth]{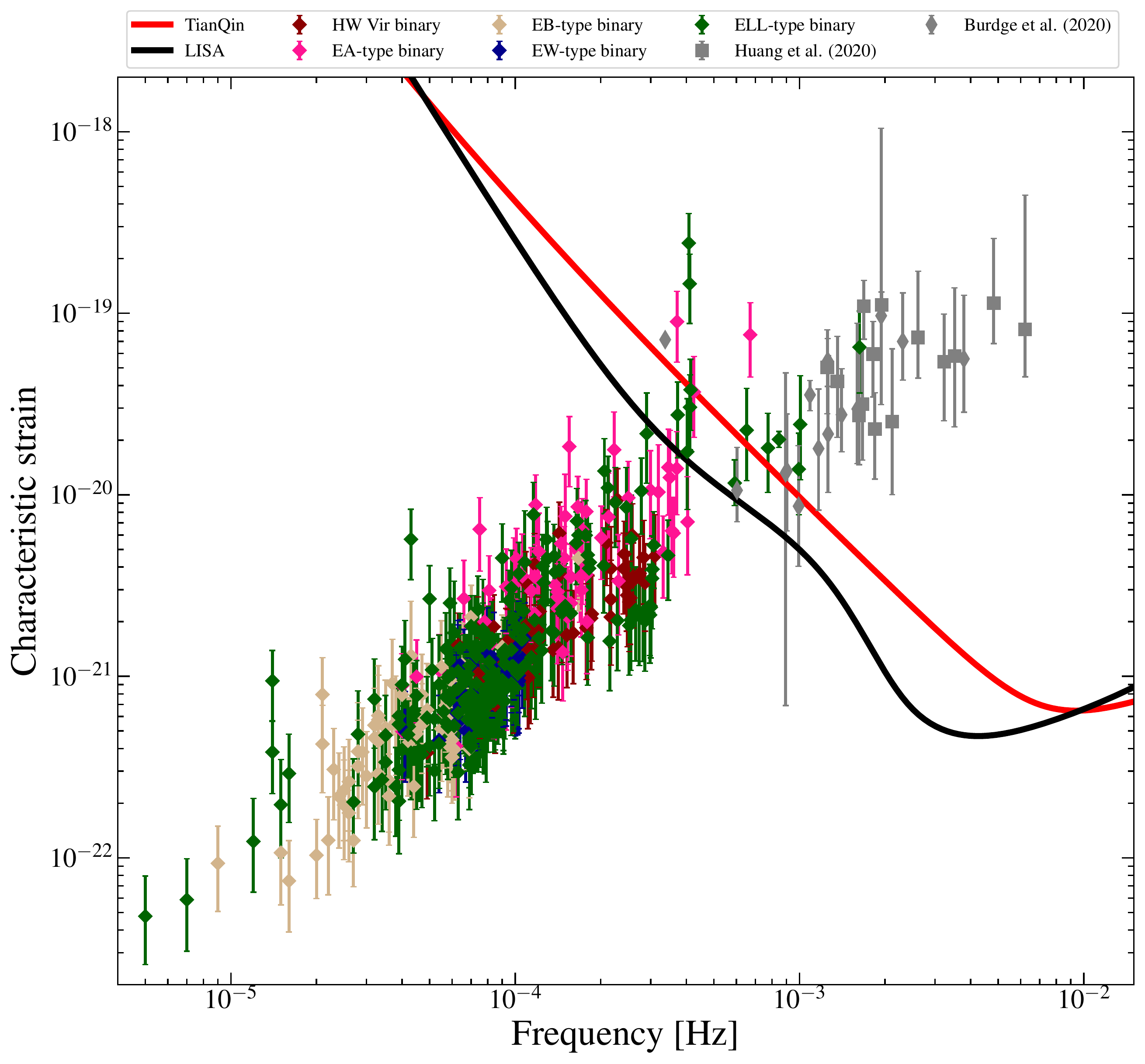}
\caption{ Characteristic gravitational-wave strain of the sources described in this work (shown as colors diamonds), compared with the samples of candidate verification binaries from \citet{Huang2020} (shown as grey squares) and \citet{Burdge2020b, Burdge2020a}, \citet{Kilic2021} and \citet{Chandra2021} (shown as grey thin diamonds), overplotted with \emph{LISA} sensitivity curve (solid black line) and \emph{TianQin} sensitivity curve (solid red line).}
\label{fig:GWhc}
\end{center}
\end{figure*}

\subsection{GW Signals of Binary Systems} \label{subsec:gwsignals}

Galactic ultracompact binaries are the most numerous \ac{GW} sources expected to be detected by future space-based \ac{GW} detectors such as TianQin and LISA \citep{Kupfer2018,Huang2020,Amaro-Seoane2022}.
The loudest signals can be detected individually, while a large number of the unresolved \ac{GW} signals will superpose to form the DWDs foreground for the space-based \ac{GW} detectors \citep{Huang2020,Liang2022,Lu2022}.
Once the short-period DWDs candidates in our final catalog are confirmed, it can help to expand the population of verification binaries. 

The \ac{GW} radiation from DWDs can be considered as quasi-monochromatic signal sources, which can be described by seven parameters: frequency $f_\mathrm {GW}= 2 / P_\mathrm {orb}$, the dimensionless amplitude $\mathcal{A}$, ecliptic coordinates ($\lambda$, $\beta$), orbital inclination $\iota$, polarization angle $\Psi_{S}$, and initial orbital phase $\phi_{0}$ \citep{Huang2020}. The primary and secondary masses ($m_{1}$ and $m_{2}$), luminosity distance ($d$), and orbital period ($P_\mathrm{orb}$) of double white dwarfs can be derived from electromagnetic (EM) observations, and can be used to estimate the amplitude of the gravitational wave signal \citep{Huang2020}:
\begin{equation}\label{eq:GWAmp}
   \mathcal{A} = \frac{2(G\mathcal{M})^{5/3}}{c^4d}(\pi f)^{2/3},
\end{equation}
where $\mathcal{M} = (m_{1}m_{2})^{3/5}(m_{1}+m_{2})^{-1/5}$, $G$ and $c$ are the chirp mass, the gravitational constant and the speed of light, respectively.

The characteristic strain can be defined by the dimensionless \ac{GW} amplitude $\mathcal{A}$ and the frequency of \ac{GW} radiation $f_\mathrm {GW}$:
\begin{equation}\label{eq:Strain}
   h_{c} = \sqrt{N_\mathrm {cycle}} \mathcal{A},
\end{equation}
where $N_\mathrm {cycle} = f_\mathrm {GW}T_\mathrm {obs}$ is the number of binary orbital cycles observed during the mission, $T_\mathrm {obs}$ is the integration time (or observation time) of the detectors.

To estimate the \ac{SNR} ($\rho$) of \ac{GW} signals from DWDs, we use the expression defined as \citet{Korol2017} and \citet{Huang2020},
\begin{equation}\label{eq:SNR}
   \rho^2=\frac{2\langle A^{2} \rangle T_{obs}}{\widetilde{S}_{n}(f_{s})},
\end{equation}
and the average amplitude defined by 
\begin{equation}\label{eq:AverageAmp}
   \left\langle A^{2} \right\rangle =\mathcal{A}^2\left[(1+\cos^2 (\iota))^2\left\langle F^2_{+} \right\rangle+4 \cos^2 (\iota) \left\langle F^2_{\times} \right\rangle \right],
\end{equation}
where $\left\langle F^2_{+}(\lambda, \beta, \Psi_{S}) \right\rangle$ and $\left\langle F^2_{\times}(\lambda, \beta, \Psi_{S}) \right\rangle$ are the orbit averaged detector responses \citep{Cornish2003, Huang2020}.
$\widetilde{S}_{n}(f_{s})$ is the sensitivity curve of the detector in a Michelson channel. The sensitivity curve of TianQin can be expressed analytically as Eqs. (13) of \citet{Huang2020}, and the sensitivity curve of the detector of LISA from \citet{Robson2019}. 

In order to estimate the SNR of binaries under the actual observation conditions of the detectors, we consider the effective observation time of the GW detectors. For TianQin, the 5-year mission adopts the observation mode of three-month observation plus three-month shutdown protection, which effectively corresponds to 2.5 years of observation time \citep{Luo2016,Huang2020}. For LISA, based on the performance of LISA Pathfinder, LISA Science Group expects that LISA will have a duty cycle of about 0.75, which means that in its 4-year mission, its effective observation time is 3 years \citep{AmaroSeoane2022}. In equation (\ref{eq:SNR}), the observation time $\rm T_{obs}$ of TianQin is set as 2.5 years, and that of LISA is set as 3 years.

In this work, we discovered 429 \ac{CWDB} candidates.
In order to calculate the \ac{SNR} of these sources, we need to obtain the source parameters. 
The sky location ($\lambda, \beta$) of sources can be used the ecliptic longitude (\texttt{ecl\_lon}) and ecliptic latitude (\texttt{ecl\_lat}) from Gaia EDR3 data.
Using BLS and CE to analyze the ZTF light curve, we can calculate the trustworthy orbital period of binary candidates.
We compared the period values obtained by the two algorithms after independently calculating three groups of ZTF photometry data, and finally estimated the average orbital period and retained it to three decimal places.
The luminosity distance can be determined using trigonometric parallaxes from Gaia EDR3 \citep{Lindegren2021,Kupfer2018}. See Table \ref{tab:CVBsources} for further details of potential GW candidates.
We assume that the primary mass $m_{1}$ obeys a uniform distribution $U(0.5,0.7)$ $\mathrm{M}_\odot$, and the secondary mass $m_{2}$ obeys a uniform distribution $U(0.2,0.4)$ $\mathrm{M}_\odot$ according to \citet{Burdge2020a}, and we fix $\iota=\pi/3$, $\Psi_S=\pi/2$ for all new candidates, since there are currently no measurements of these parameters.

We report the estimated TianQin/LISA SNR for the sources in the sample in Table \ref{tab:CVBsources}.
The first part of the Table \ref{tab:CVBsources} shows the first 24 candidates with periods less than 100 minutes in our binary sample, which are defined as potential GW candidates (for further details on their light curves, please see Appendix \ref{ref:appendixd}). The later part of the Table \ref{tab:CVBsources} shows the recently discovered GW sources by other surveys that is not summarised in \citet{Huang2020}.
Figure \ref{fig:GWhc} shows the characteristic strain of gravitational radiation of our newly discovered binary samples and other classic verification binaries and the sensitivity of TianQin and LISA.
As illustrated in Figure \ref{fig:GWhc}, most sources of our sample fall below the sensitivity curves of TianQin and LISA, about 10 candidates fall above the LISA sensitivity curve, and about 16 candidates fall above the LISA sensitivity curve after 4 years of observations. In estimating the GW signals of candidates, we fixed some source parameters as constants, and the mass was uniformly distributed (we remark that for some binary systems this assumption might be unphysical), the current uncertainty in the SNR and GW amplitude of all sources originate from the uncertainty of distance measurement (mainly due to the standard error of parallax, \texttt{parallax\_error}). 
In our binary sample, the light curve shapes of potential GW candidates located above or near the sensitivity curve are mainly EA-type and ELL-type binaries.
We adopt a low SNR threshold of 5 as the minimum standard for GW signals detected by TianQin and LISA. 
For TianQin, we found two new candidates VBs, namely ZTF J0526+5934 (SNR $\sim 12$) and ZTF J2007+1742 (SNR $\sim 7$), plus other newly discovered GW sources in the later part of Table \ref{tab:CVBsources}, bringing the total number of VBs to 18 for TianQin.
For LISA, we found 6 new VBs candidates, bringing the total number of VBs to 31 for LISA.

\begin{deluxetable*}{cccccccc}[htpb]
\tablenum{3}
\tablecaption{The sources parameter, amplitude $\mathcal{A}$ and \ac{SNR} for 24 ultra-short period ($P_{\rm orb} \leq$ 100~min) potential GW candidates. \label{tab:CVBsources}}
\tablewidth{0pt}
\tablehead{
\colhead{ZTF Name} & \colhead{$\lambda$} & \colhead{$\beta$} & \colhead{$f$} & \colhead{$d$}  &  \colhead{$\mathcal{A}$}  & \colhead{SNR}  & \colhead{SNR} \\
\colhead{} & \colhead{[deg]} & \colhead{[deg]} & \colhead{[mHz]} & \colhead{[kpc]}  & \colhead{[$\times 10^{-23}$]} & \colhead{[TianQin]} & \colhead{[LISA]}
}
\startdata
ZTF J0526+5934 & 84.6989 & 36.2955 & 1.626 & 0.846 & 18.15 & 12.777 & 35.788 \\
ZTF J1840-1742 & 279.7729 & 5.3797 & 1.006 & 1.345 & 8.66 & 3.297 & 5.439 \\
ZTF J1707-1522 & 257.2883 & 7.4854 & 0.994 & 2.285 & 4.94 & 1.477 & 2.931 \\
ZTF J2130+4420$^{[b]}$ & 348.1685 & 54.4443 & 0.847 & 1.309 & 7.81 & 0.850 & 2.076  \\
ZTF J2351+6305 & 36.8445 & 55.5895 & 0.775 & 1.286 & 7.32 & 0.681 & 2.409  \\
ZTF J1611+6308 & 183.9169 & 78.0916 & 0.671 & 0.257 & 33.07 & 2.282 & 8.147 \\
ZTF J1813+4251$^{[e]}$ & 276.0031 & 66.2346 & 0.652 & 0.835 & 10.00 & 0.901 & 2.415 \\
ZTF J2055+4651$^{[b]}$ & 341.0544 & 59.9596 & 0.592 & 2.264 & 5.37 &  0.283 & 0.779  \\
ZTF J1603+2150 & 232.8645 & 41.6031 & 0.425 & 0.309 & 20.11 & 0.667 & 2.489 \\
ZTF J0451+0104 &  71.5240 & -21.2702 & 0.414 & 0.294 & 20.96 & 0.924 & 2.568 \\
ZTF J0112+5827 & 44.5245 & 45.8154 & 0.412 & 0.364 & 16.83 & 0.433 & 1.915 \\
ZTF J2334+3921 & 12.6230 &  38.0721 & 0.411 & 0.076 & 80.71 & 2.374 & 9.624 \\
ZTF J2007+1742 & 309.0298 & 36.9204 & 0.408 & 0.045 & 135.70 & 7.219 & 13.969 \\
ZTF J1810-2138 & 272.3779 & 1.7778 & 0.404 & 1.683 & 3.97 & 0.217 & 0.513 \\
ZTF J1705+3455 & 249.0813 & 57.3809 & 0.403 & 0.616 & 9.68 & 0.309 & 1.047 \\
ZTF J1943-2657 & 293.0642 & -5.5741 & 0.373 & 0.358 & 16.05 & 0.814 & 1.615 \\
ZTF J1708-2548 & 258.3723 & -2.8898 & 0.371 & 0.109 & 52.53 & 2.042 & 5.605 \\
ZTF J1948+5250 & 327.3373 & 70.9334 & 0.370 & 0.699 & 8.17 &  0.208 & 0.721 \\
ZTF J1944+5449 & 329.5948 & 72.8635 & 0.361 & 1.552 & 3.64 & 0.085 & 0.306 \\
ZTF J1927+3411 & 302.7131 & 55.1734 & 0.356 & 1.482 & 3.78 & 0.121 & 0.320 \\
ZTF J0609+3652 & 91.8999 & 13.4434 & 0.350 & 0.733 & 7.49 & 0.288 & 0.705 \\
ZTF J1100+5210 & 142.3087 & 41.4823 & 0.346 & 0.636 & 8.57 & 0.259 & 0.655 \\
ZTF J0050+2551 & 21.9928 & 18.7691 & 0.345 & 1.963 & 2.81 & 0.051 & 0.242 \\
ZTF J1622+4730 & 224.3785 & 67.1495 & 0.332 & 1.814 & 2.98 & 0.050 & 0.212 \\
\hline 
ZTF J2243+5242$^{[a]}$ & 13.2423  &  53.9599 & 3.788 & 1.753 & 10.26 &  19.992 & 78.739  \\
ZTF J0538+1953$^{[b]}$  & 84.8061  &  -3.4356 & 2.308 & 0.997 & 16.38 &  18.915 & 75.532  \\
ZTF J1905+3134$^{[b]}$  & 293.7825 &  53.6335 & 1.938 & 0.696 & 24.78 &  26.305 & 79.343  \\
ZTF J2029+1534$^{[b]}$  & 314.4631 &  33.4205 & 1.597 & 1.095 &  8.36 &   4.743 & 9.144  \\
ZTF J0722-1839$^{[b]}$  & 115.8782 & -40.2164 & 1.406 & 1.267 &  8.29 &   3.451 & 6.666  \\
ZTF J1749+0924$^{[b]}$  & 267.0209 &  32.8576 & 1.261 & 1.291 &  6.85 &   2.178 & 4.685  \\
ZTF J2228+4949$^{[b]}$  & 7.0683   &  53.2065 & 1.167 & 2.076 &  5.92 &   1.520 & 4.416  \\
ZTF J1946+3203$^{[b]}$  & 307.9775 &  52.0541 & 0.993 & 1.919 &  3.09 &   0.626 & 1.186  \\
ZTF J0643+0318$^{[b]}$  & 101.5702 & -19.6948 & 0.903 & 2.040 &  5.08 &   1.519 & 2.398  \\
ZTF J0640+1738$^{[b]}$  & 99.6432  &  -5.4567 & 0.894 & 1.576 &  4.97 &   1.339 & 2.262  \\
ZTF J2320+3750$^{[b]}$  & 8.7171   &  38.0936 & 0.603 & 1.256 &  4.85 &   0.244 & 0.793  \\
SDSS J0634+3803$^{[c]}$ & 97.0832  &  14.8390 & 1.257 & 0.433 & 17.36 &  13.782 & 23.919  \\
SMSS J0338-8139$^{[c]}$ & 286.4357 & -72.7068 & 1.089 & 0.536 & 12.11 &   2.199 & 6.226  \\
SDSS J1337+3952$^{[d]}$ & 182.8931 &  45.5716 & 0.337 & 0.114 & 43.86 &   1.151 & 5.030  \\
\enddata
\tablecomments{The position ($\lambda$, $\beta$) are the ecliptic coordinates, $f$ is the \ac{GW} frequency, $\mathcal{A}$ is the dimensionless amplitude, $d$ is the luminosity distance. 
\tablerefs{\citet{Burdge2020b}$^{[a]}$,\citet{Burdge2020a}$^{[b]}$,\citet{Kilic2021}$^{[c]}$,\citet{Chandra2021}$^{[d]}$,\citet{Burdge2022}$^{[e]}$}}
\end{deluxetable*}

\section{Conclusion} \label{sec:conclusion}

In this work, we have presented a catalog of short-period close white dwarf binary candidates based on the Gaia EDR3 catalog and the Zwicky Transient Facility photometry data. We defined a color-cutting criterion for selecting initial samples in the Gaia H-R diagram. In the Gaia EDR3 data, we searched 823231 high signal-to-noise ratio sources after using quality filtering parameters \citep{Pelisoli2019b, Geier2019}. We applied the Gaia variability metric \citep{Guidry2021} to select 12480 most variable objects with high confidence from high-SNR samples. 

We cross-matched the high-confidence variable source catalog with time-domain photometry data of the ZTF Public DR8 and analyzed the light curves of all variable sources. After analyzing the ZTF light curves, a total of 826 candidates were identified to have distinguishable periodic variability. Taking the shape of the light curve produced by the eclipse process of the binary system as the selection criterion, we found 429 binary candidates. The final catalog includes 58 HW Vir-type binaries, 65 EA-type binaries, 56 EB-type binaries, 41 EW-type binaries, and 209 ELL-type binaries.

We analyzed four short-period close white dwarf binary candidates with special light curves in the final catalog. 
ZTF J2351+6305 is an ellipsoidal variable binary with a period of less than 60 minutes ($\mathrm P_\mathrm {orb}$ = 42.985 min), which is similar to a mass-transferring and recently-detached CV \citep{El-Badry2021b}. 
ZTF J1611+6308 is a deeply-eclipsing white dwarf binary with an orbital period of 49.708 minutes. 
ZTF J1813+4251 is a short-period ($\mathrm P_\mathrm {orb}$ = 51.161 min) binary candidate near the classical-CV in the Gaia H-R diagram. 
ZTF J0112+5827 is a CV candidate similar to an eclipsing polar BS tri system \citep{Kolbin2022} with a period of 80.912 minutes. 
We obtained the ZTF photometric data of these special candidates and analyzed their positions in the Gaia H-R diagram. The future follow-up observation of the spectra can be used to estimate the physical parameters such as masses, temperatures, and radial velocity semiamplitudes. The high-speed photometric observation can be used to solve the orbital parameters of the binary system, such as radius, inclination, semimajor axis, and orbital period.

We define the distribution of the Gold samples of close-WD binaries based on the sub-classes of stars in the Gaia H-R diagram. In the final catalog, a total of 429 close-WD binary candidates were divided into 85 PCEB Gold samples, 224 CV Gold samples, 45 ELM-WD Gold samples, and 11 DWD Gold samples. 

We estimated the gravitational wave amplitudes and signal-to-noise ratios of all candidates in our binary sample.  We found that we have two potential GW candidates with SNRs greater than 5 in the 2.5-year observation time of TianQin, which increases the total number of candidate VBs for TianQin to 18. For LISA, we have six new sources with SNRs of more than 5 in their 3-year effective observation time with a total of 4-year mission lifetime. The total number of LISA VBs has reached 31.

In future work, we aim to use multi-band spectral energy distribution to analyze the compositions and obtain the effective temperature and surface gravity of the binary systems to study their evolution process. We also plan to use LAMOST spectral and jointly fit the light curve to obtain the physical parameters and orbital parameters for our binary candidates to confirm their binary nature.

\acknowledgments
We thank Jian-dong Zhang, En-Kun Li, and Siyi Xu for their helpful comments and discussions.
This work has been supported by the Guangdong Major Project of Basic and Applied Basic Research (Grant No. 2019B030302001), the National Natural Science Foundation of China (Grant No. 12233013, No. 12173104), and the National Key Research and Development Program of China (Grant No. 2020YFC2201400). B. M. acknowledges financial support from the National Natural Science Foundation of China (grant Nos. 12073092, 12103097) and the science research grants from the China Manned Space Project (CMS-CSST-2021-B09，CMS-CSST- 2021-B12).

This work has made use of data from the European Space Agency (ESA) mission Gaia (\url{https://www.cosmos.esa.int/gaia}), processed by the Gaia Data Processing and Analysis Consortium (DPAC, \url{https://www.cosmos.esa.int/web/gaia/dpac/consortium}). Funding for the DPAC has been provided by national institutions, in particular the institutions participating in the Gaia Multilateral Agreement. 
The ZTF data used in this study can be accessed from the ZTF Science Data System (ZSDS) \citep{Masci2019} housed at NASA/IPAC Infrared Science Archive (IRSA) (IRSA ZTF Lightcurve Queries
 API, \url{https://irsa.ipac.caltech.edu/docs/program_interface/ztf_lightcurve_api.html}).


\facility{Gaia, PO:1.2 m (ZTF), ADS, CDS.}
\software{ Astropy \citep{Astropy2013}, matplotlib \citep{Hunter2007}, NumPy \citep{vanderWalt2011,Harris2020}, Pandas \citep{McKinney2010}, SciPy \citep{Virtanen2020}, the IPython package \citep{Perez2007}, and the Time Series Tool \citep{https://doi.org/10.26131/irsa538}}.

\bibliography{sample1}{}
\bibliographystyle{aasjournal}

\appendix

\section{Gaia astrometric query cut} \label{ref:appendixa}

Below we show the initial-cut selection criteria applied to Gaia EDR3\footnote{\url{https://gea.esac.esa.int/archive/}} data through the following Astronomical Data Query Language (ADQL) query. Our ADQL query jobs are run in the Gaia@AIP\footnote{\url{https://gaia.aip.de/}} services or ARI's Gaia\footnote{\url{https://gaia.ari.uni-heidelberg.de/tap.html}} services.

\begin{lstlisting}
SELECT * 
FROM gaiaedr3.gaia_source 
WHERE dec > -28.0
AND bp_rp < 1.0 
AND parallax_over_error > 5 
AND phot_bp_mean_flux_over_error > 10 
AND phot_rp_mean_flux_over_error > 10 
AND 5 + 5 * log10((parallax + 0.029)/1000) + phot_g_mean_mag > 3.7 * bp_rp + 2.2 
AND 5 + 5 * log10((parallax + 0.029)/1000) + phot_g_mean_mag < 2.7 * bp_rp + 11.5 
AND phot_bp_rp_excess_factor < 1.45 + 0.06 * power(phot_bp_mean_mag - phot_rp_mean_mag, 2) 
AND phot_bp_rp_excess_factor > 1.0 + 0.015 * power(phot_bp_mean_mag - phot_rp_mean_mag, 2) 
AND (astrometric_chi2_al / (astrometric_n_good_obs_al - 5) < 1.44 
OR astrometric_chi2_al / (astrometric_n_good_obs_al - 5) < 1.44 * exp(- 0.4 * (phot_g_mean_mag - 19.5)));
\end{lstlisting}

Below we show the Gaia Variability cut selection criteria using the following ADQL query:

\begin{lstlisting}
SELECT * 
FROM gaiaedr3.gaia_source 
WHERE dec > -28.0
AND bp_rp < 1.0 
AND parallax_over_error > 5 
AND phot_bp_mean_flux_over_error > 10 
AND phot_rp_mean_flux_over_error > 10 
AND 5 + 5 * log10((parallax + 0.029)/1000) + phot_g_mean_mag > 3.7 * bp_rp + 2.2 
AND 5 + 5 * log10((parallax + 0.029)/1000) + phot_g_mean_mag < 2.7 * bp_rp + 11.5 
AND phot_bp_rp_excess_factor < 1.45 + 0.06 * power(phot_bp_mean_mag - phot_rp_mean_mag, 2) 
AND phot_bp_rp_excess_factor > 1.0 + 0.015 * power(phot_bp_mean_mag - phot_rp_mean_mag, 2) 
AND (astrometric_chi2_al / (astrometric_n_good_obs_al - 5) < 1.44 
OR astrometric_chi2_al / (astrometric_n_good_obs_al - 5) < 1.44 * exp(- 0.4 * (phot_g_mean_mag - 19.5)))
AND phot_g_mean_flux_error * power(phot_g_n_obs, 0.5) / phot_g_mean_flux - ((8.31e-9) * exp(0.794 * phot_g_mean_mag) + 0.0005 * exp(phot_g_mean_mag - 17.0) + 0.019) > 0.0;
\end{lstlisting}

\section{Short-period CWDBs catalogs} \label{ref:appendixb}

Below we show the Gaia photometric parameters and the orbital period parameters for all 429 objects in our binary sample. According to the types of the light curve, we can divide them into five tables: HW Vir-type binary candidates (see Table \ref{tab:HWVirsources}), EA-type binary candidates (see Table \ref{tab:EAsources}), EB-type binary candidates (see Table \ref{tab:EBsources}), EW-type binary candidates (see Table \ref{tab:EWsources}), and ELL-type binary candidates (see Table \ref{tab:ELLsources}).

\startlongtable
\begin{deluxetable}{cccccccccc}
\tablenum{4}
\tablecaption{HW Vir-type binary candidates \label{tab:HWVirsources}}
\tablewidth{0pt}
\tablehead{
\colhead{ZTF Name} & \colhead{Gaia EDR3} &  \colhead{RA}  & \colhead{Dec} & \colhead{$\varpi$} & \colhead{G} & \colhead{$G_\mathrm{BP}-G_\mathrm{RP}$} & \colhead{$G_\mathrm{abs}$} &
\colhead{$P_\mathrm{orb}$}  & \colhead{$\mathcal{A}$} \\
\colhead{} & \colhead{sourceID} & \colhead{(deg)} & \colhead{(deg)} & \colhead{(mas)} & \colhead{(mag)} &  \colhead{(mag)} & \colhead{(mag)} & \colhead{(min)}  & \colhead{($\times 10^{-23}$)}
}
\startdata
ZTF J2040+3407 & 1864121982365984128 & 310.19402 & 34.11743  & 0.55 & 17.41 & 0.42 & 6.14 & 107.543 & 2.94 \\
ZTF J2345+5057 & 1943858286426802304 & 356.26684 & 50.96520  & 0.43 & 16.43 & -0.13 & 4.62 & 116.686 & 2.18 \\
ZTF J2120+4014 & 1965596391525733120 & 320.17691 & 40.24297  & 0.62 & 17.02 & 0.25 & 5.99 & 117.598 & 3.02 \\
ZTF J0718+0739 & 3154801672464733440 & 109.74027 & 7.65367   & 0.52 & 15.22 & -0.41 & 3.82 & 121.625 & 2.51 \\
ZTF J0836-0802 & 5753841281270433536 & 129.05426 & -8.03995  & 0.51 & 15.28 & -0.37 & 3.81 & 127.971 & 2.35 \\
ZTF J2328+6415 & 2209063896962507520 & 352.22684 & 64.26275  & 0.53 & 17.37 & 0.91 & 5.99 & 128.107 & 2.47 \\
ZTF J2046+5147 & 2182023160826160000 & 311.65906 & 51.79319  & 0.93 & 15.22 & 0.18 & 5.07 & 129.086 & 4.19 \\
ZTF J0518-2308 & 2961366588952039552 & 79.52903 & -23.14588 & 0.42 & 15.89 & -0.35 & 3.97 & 131.287 & 1.91 \\
\enddata
\tablecomments{The coordinates and basic photometric characteristics of the 58 HW Vir-type binary candidates discovered so far using ZTF data. Coordinates are taken from \emph{Gaia} and are in J2000.0. The IAU name is provided in the catalog.
\tablerefs{\citet{Kao2016}$^{[1]}$,\citet{Geier2011}$^{[2]}$,\citet{Almeida2012}$^{[3]}$,\citet{Keller2022}$^{[4]}$,\citet{Geier2017}$^{[5]}$,\citet{For2010}$^{[6]}$}}
\end{deluxetable}
\onecolumngrid

\startlongtable
\begin{deluxetable}{cccccccccc}
\tablenum{5}
\tablecaption{EA-type binary candidates \label{tab:EAsources}}
\tablewidth{0pt}
\tablehead{
\colhead{ZTF Name} & \colhead{Gaia EDR3} &  \colhead{RA}  & \colhead{Dec} & \colhead{$\varpi$} & \colhead{G} & \colhead{$G_\mathrm{BP}-G_\mathrm{RP}$} & \colhead{$G_\mathrm{abs}$} &
\colhead{$P_\mathrm{orb}$}  & \colhead{$\mathcal{A}$} \\
\colhead{} & \colhead{sourceID} & \colhead{(deg)} & \colhead{(deg)} & \colhead{(mas)} & \colhead{(mag)} &  \colhead{(mag)} & \colhead{(mag)} & \colhead{(min)}  & \colhead{($\times 10^{-23}$)}
}
\startdata
ZTF J1611+6308 & 1629388752470472704 & 242.89152 & 63.14213  & 3.89 & 18.28 & 0.19 & 11.24 & 49.708 & 33.07 \\
ZTF J1603+2150 & 1206052590817661696 & 240.89972 & 21.84231  & 3.23 & 19.11 & 0.19 & 11.66 & 78.394 & 20.11 \\
ZTF J1810-2138 & 4093796151752232192 & 272.55704 & -21.64038 & 0.62 & 17.54 & 0.72 & 6.51 & 82.426 & 3.97 \\
ZTF J1708-2548 & 4111991385628196224 & 257.07942 & -25.80907 & 9.21 & 15.36 & 0.62 & 10.19 & 89.896 & 52.53 \\
ZTF J1948+5250 & 2137193078157620480 & 297.08732 & 52.84653  & 1.43 & 18.50 & 0.53 & 9.28 & 90.134 & 8.17 \\
ZTF J1944+5449 & 2138663782338254464 & 296.17858 & 54.82857  & 0.64 & 15.75 & -0.23 & 4.79 & 92.393 & 3.64 \\
ZTF J1927+3411 & 2046661975094606208 & 291.90956 & 34.19492  & 0.67 & 15.26 & -0.31 & 4.41 & 93.522 & 3.78 \\
ZTF J0609+3652$^{[1]}$ & 3453995450857025664 & 92.30994 & 36.86744  & 1.36 & 18.79 & -0.03 & 9.47 & 95.105 & 7.49 \\
\enddata
\tablecomments{The coordinates and basic photometric characteristics of the 65 EA-type binary candidates discovered so far using ZTF data. Coordinates are taken from \emph{Gaia} and are in J2000.0. The IAU name is provided in the catalog.
\tablerefs{\citet{Keller2022}$^{[1]}$,\citet{Kosakowski2022}$^{[2]}$,\citet{Barlow2013}$^{[3]}$,\citet{Jiang2013}$^{[4]}$,\citet{Guo2015}$^{[5]}$,\citet{Dimitrov2012}$^{[6]}$,\citet{Kjurkchieva2015}$^{[7]}$}}
\end{deluxetable}
\onecolumngrid

\startlongtable
\begin{deluxetable}{cccccccccc}
\tablenum{6}
\tablecaption{EB-type binary candidates \label{tab:EBsources}}
\tablewidth{0pt}
\tablehead{
\colhead{ZTF Name} & \colhead{Gaia EDR3} &  \colhead{RA}  & \colhead{Dec} & \colhead{$\varpi$} & \colhead{G} & \colhead{$G_\mathrm{BP}-G_\mathrm{RP}$} & \colhead{$G_\mathrm{abs}$} &
\colhead{$P_\mathrm{orb}$}  & \colhead{$\mathcal{A}$} \\
\colhead{} & \colhead{sourceID} & \colhead{(deg)} & \colhead{(deg)} & \colhead{(mas)} & \colhead{(mag)} &  \colhead{(mag)} & \colhead{(mag)} & \colhead{(min)}  & \colhead{($\times 10^{-23}$)}
}
\startdata
ZTF J2044+2653 & 1856208999031665664 & 311.19562 & 26.88862 & 1.15 & 18.74 & 0.69 & 9.05 & 199.497 & 3.90 \\
ZTF J1301+1631 & 3937384472308048256 & 195.44488 & 16.52939 & 0.77 & 14.52 & 0.47 & 3.95 & 369.066 & 1.72 \\
ZTF J0515+0104 & 3233813338246568448 & 78.92258  & 1.07789 & 1.52 & 15.73 & 0.97 & 6.65 & 478.965 & 2.86 \\
ZTF J1716+7325 & 1655226867702084352 & 259.09056 & 73.42544 & 0.25 & 16.39 & 0.32 & 3.39 & 479.665 & 0.54 \\
ZTF J0101-1952 & 2356059083939929344 & 15.39862 & -19.87515 & 0.40 & 15.49 & 0.36 & 3.49 & 494.834 & 0.75 \\
ZTF J0734-0406 & 3057327664560134144 & 113.65574 & -4.11209 & 0.89 & 16.89 & 0.91 & 6.64 & 511.579 & 1.62 \\
ZTF J2241+4639 & 1983499150940152064 & 340.35358 & 46.65429 & 0.73 & 17.94 & 0.81 & 7.24 & 542.995 & 1.31 \\
ZTF J0324-0715 & 5168804986809073280 & 51.16995 & -7.26322 & 0.29 & 16.23 & 0.29 & 3.54 & 548.786 & 0.57 \\
\enddata
\tablecomments{The coordinates and basic photometric characteristics of the 56 EB-type binary candidates discovered so far using ZTF data. Coordinates are taken from \emph{Gaia} and are in J2000.0. The IAU name is provided in the catalog.}
\end{deluxetable}
\onecolumngrid

\startlongtable
\begin{deluxetable}{cccccccccc}
\tablenum{7}
\tablecaption{EW-type binary candidates \label{tab:EWsources}}
\tablewidth{0pt}
\tablehead{
\colhead{ZTF Name} & \colhead{Gaia EDR3} &  \colhead{RA}  & \colhead{Dec} & \colhead{$\varpi$} & \colhead{G} & \colhead{$G_\mathrm{BP}-G_\mathrm{RP}$} & \colhead{$G_\mathrm{abs}$} &
\colhead{$P_\mathrm{orb}$}  & \colhead{$\mathcal{A}$} \\
\colhead{} & \colhead{sourceID} & \colhead{(deg)} & \colhead{(deg)} & \colhead{(mas)} & \colhead{(mag)} &  \colhead{(mag)} & \colhead{(mag)} & \colhead{(min)}  & \colhead{($\times 10^{-23}$)}
}
\startdata
ZTF J0825+4329 & 916342443812848000  & 126.31920 &  43.49201 & 0.69 & 16.08 & 0.86 & 5.29 & 309.922 & 1.76 \\
ZTF J1236+2834 & 4010630775855198848 & 189.08203 &  28.56893 & 0.52 & 16.77 & 0.68 & 5.37 & 314.220 & 1.37 \\
ZTF J1141+5713 & 845729741196604544  & 175.40832 &  57.22566 & 0.41 & 17.71 & 0.97 & 5.77 & 322.677 & 1.12 \\
ZTF J0331+0207 & 3268078209375996288  & 52.96629 &   2.11906 & 0.38 & 17.04 & 0.77 & 4.97 & 323.731 & 1.05 \\
ZTF J0929+3410 & 797507570680645888  & 142.26295 &  34.16816 & 0.66 & 17.05 & 0.95 & 6.14 & 324.423 & 1.66 \\
ZTF J1452-0117 & 3650361600441639040 & 223.12124 &  -1.28868 & 0.52 & 17.09 & 0.89 & 5.68 & 324.529 & 1.34 \\
ZTF J1007-2331 & 5665664881129095040 & 151.79161 & -23.52445 & 0.38 & 16.55 & 0.53 & 4.45 & 325.688 & 1.00 \\
ZTF J2203+1624 & 1775594696266837888 & 330.90915 &  16.40823 & 0.39 & 16.82 & 0.72 & 4.78 & 333.263 & 1.03 \\
\enddata
\tablecomments{The coordinates and basic photometric characteristics of the 41 EW-type binary binary candidates discovered so far using ZTF data. Coordinates are taken from \emph{Gaia} and are in J2000.0. The IAU name is provided in the catalog.
\tablerefs{The Open Cataclysmic Variable Catalog$^{[1]}$(\url{https://depts.washington.edu/catvar/index.html})}}
\end{deluxetable}
\onecolumngrid

\startlongtable
\begin{deluxetable}{cccccccccc}
\tablenum{8}
\tablecaption{ELL-type binary candidates \label{tab:ELLsources}}
\tablewidth{0pt}
\tablehead{
\colhead{ZTF Name} & \colhead{Gaia EDR3} &  \colhead{RA}  & \colhead{Dec} & \colhead{$\varpi$} & \colhead{G} & \colhead{$G_\mathrm{BP}-G_\mathrm{RP}$} & \colhead{$G_\mathrm{abs}$} &
\colhead{$P_\mathrm{orb}$}  & \colhead{$\mathcal{A}$} \\
\colhead{} & \colhead{sourceID} & \colhead{(deg)} & \colhead{(deg)} & \colhead{(mas)} & \colhead{(mag)} &  \colhead{(mag)} & \colhead{(mag)} & \colhead{(min)}  & \colhead{($\times 10^{-23}$)}
}
\startdata
ZTF J0526+5934 & 282679289838317184   & 81.54342 &  59.57926 & 1.18 & 17.56 & 0.19 & 7.93 & 20.506 & 18.15 \\
ZTF J1840-1742 & 4099443479729561472 & 280.21873 & -17.71363 & 0.74 & 18.13 & 0.26 & 7.49 & 33.145 & 8.66 \\
ZTF J1707-1522 & 4139067134284899456 & 256.92227 & -15.37861 & 0.44 & 15.42 & 0.25 & 3.63 & 33.539 & 4.94 \\
ZTF J2130+4420$^{[1]}$ & 1970951356755394688 & 322.73629 & 44.34623 & 0.76 & 15.44 & -0.04 & 4.85 & 39.340 & 7.81 \\
ZTF J2351+6305 & 2016087511661985280 & 357.81413 &  63.09103 & 0.78 & 16.70 & 0.58 & 6.16 & 42.985 & 7.32 \\
ZTF J1813+4251 & 2113285228603943168 & 273.29631 & 42.86401 & 1.19 & 18.92 & 0.79 & 9.31 & 51.161 & 10.00 \\
ZTF J2055+4651$^{[1]}$ & 2163801981105229184 & 313.81657 & 46.85177 & 0.44 & 17.61 & 0.34 & 5.83 & 56.348 & 5.37 \\
ZTF J0451+0104 & 3231653210215464832  & 72.82015 &   1.07395 & 3.41 & 15.34 & 0.52 & 8.01 & 80.522 & 20.96 \\
\enddata
\tablecomments{The coordinates and basic photometric characteristics of the 209 ELL-type binary binary candidates discovered so far using ZTF data. Coordinates are taken from \emph{Gaia} and are in J2000.0. The IAU name is provided in the catalog.
\tablerefs{\citet{El-Badry2021b}$^{[1]}$,\citet{Barlow2013}$^{[2]}$, The Open Cataclysmic Variable Catalog$^{[3]}$(\url{https://depts.washington.edu/catvar/index.html})}}
\end{deluxetable}
\onecolumngrid

\section{Frequency spectrum} \label{ref:appendixc}

\begin{figure*}[htpb]
\begin{center}
\includegraphics[width=1.0\textwidth]{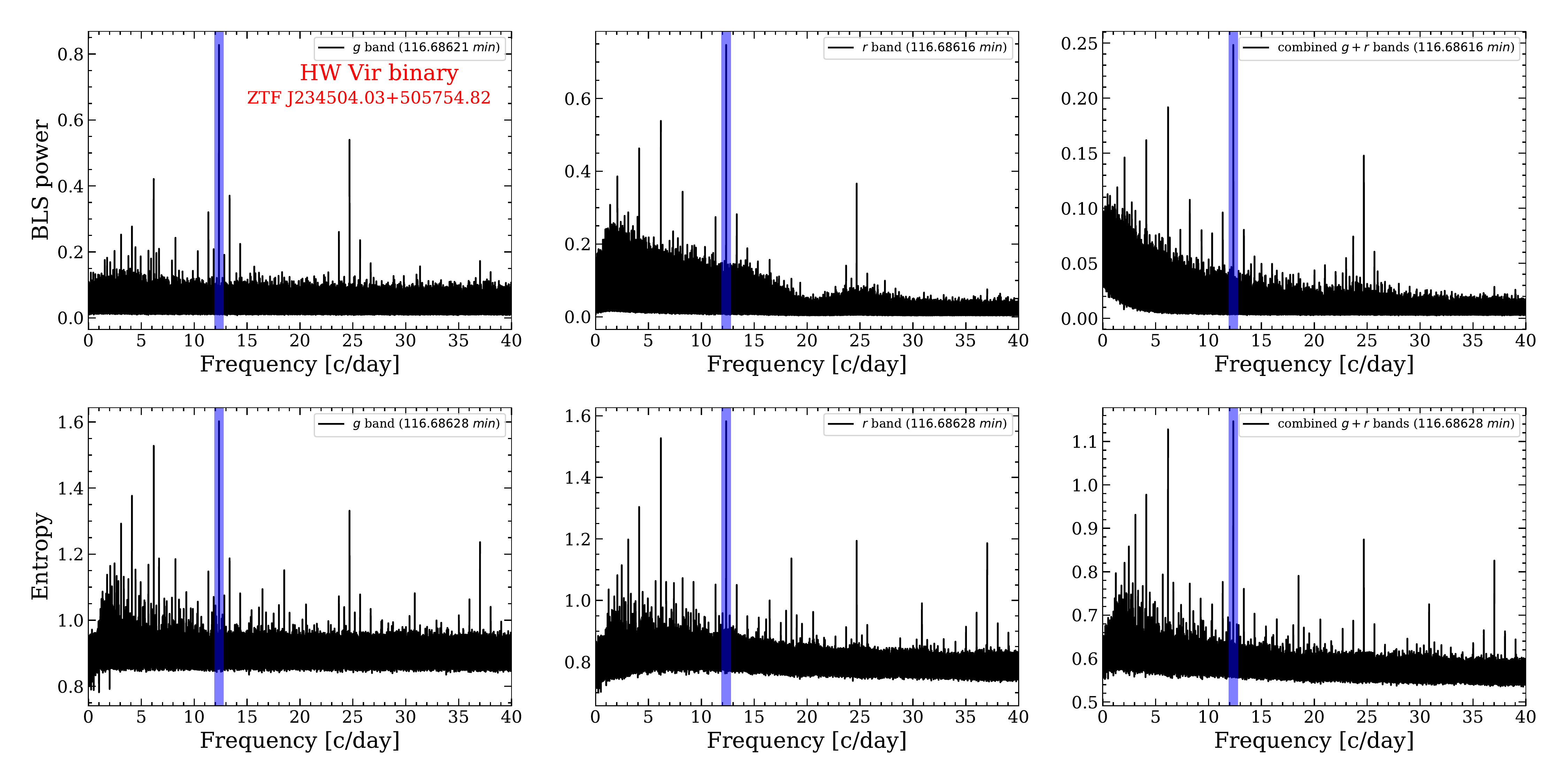}
\includegraphics[width=1.0\textwidth]{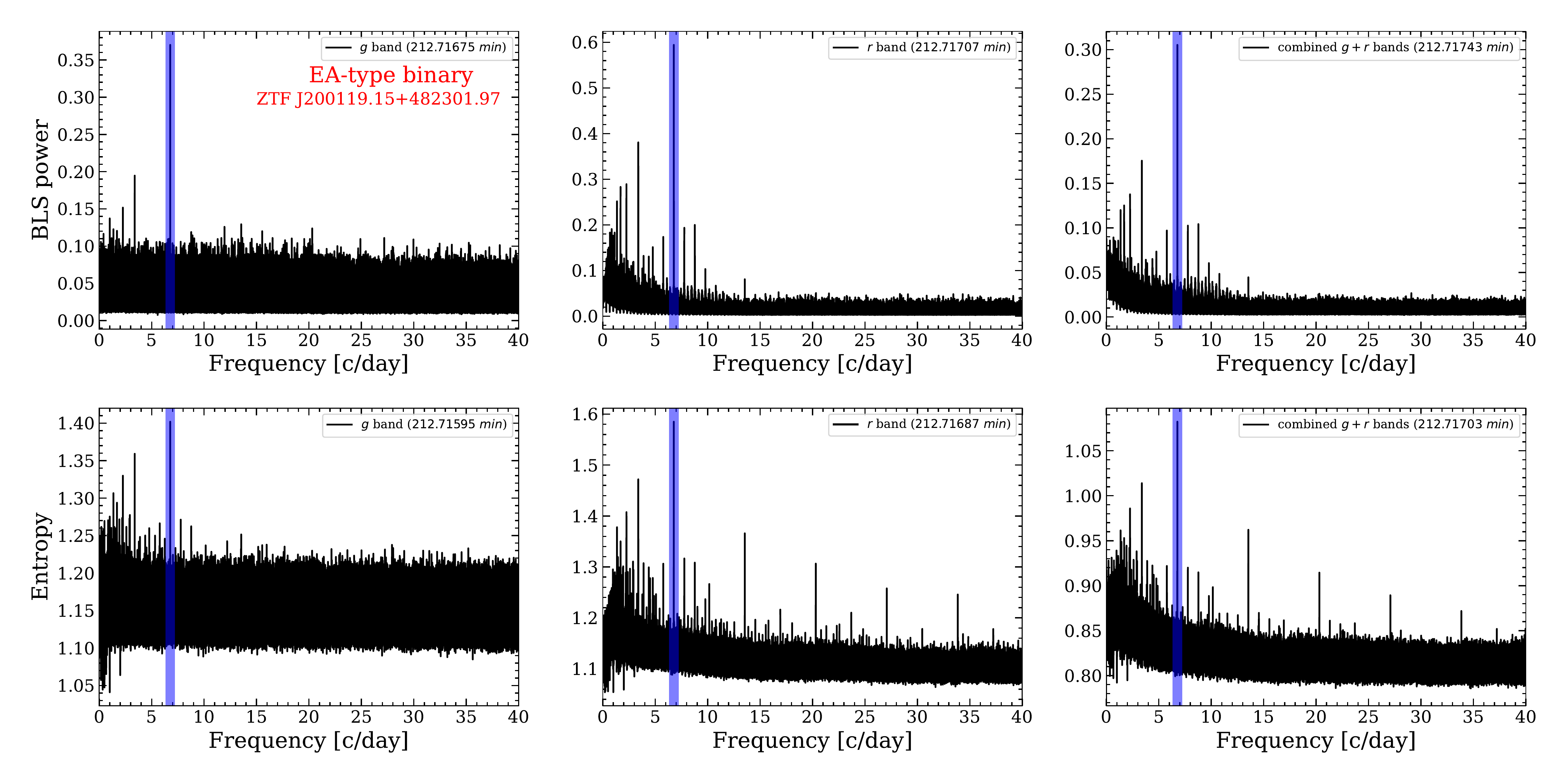}
\caption{Examples of the variability types of light curve shapes caused by the binary eclipse effect in our binary sample. All panels show the frequency spectra of the BLS and CE corresponding to the above (Figure~\ref{fig:LCexample}) variability types of the HW Vir binary and EA-type binary.}
\label{fig:BLSCEFS1}
\end{center}
\end{figure*}

\begin{figure*}[htpb]
\begin{center}
\includegraphics[width=1.0\textwidth]{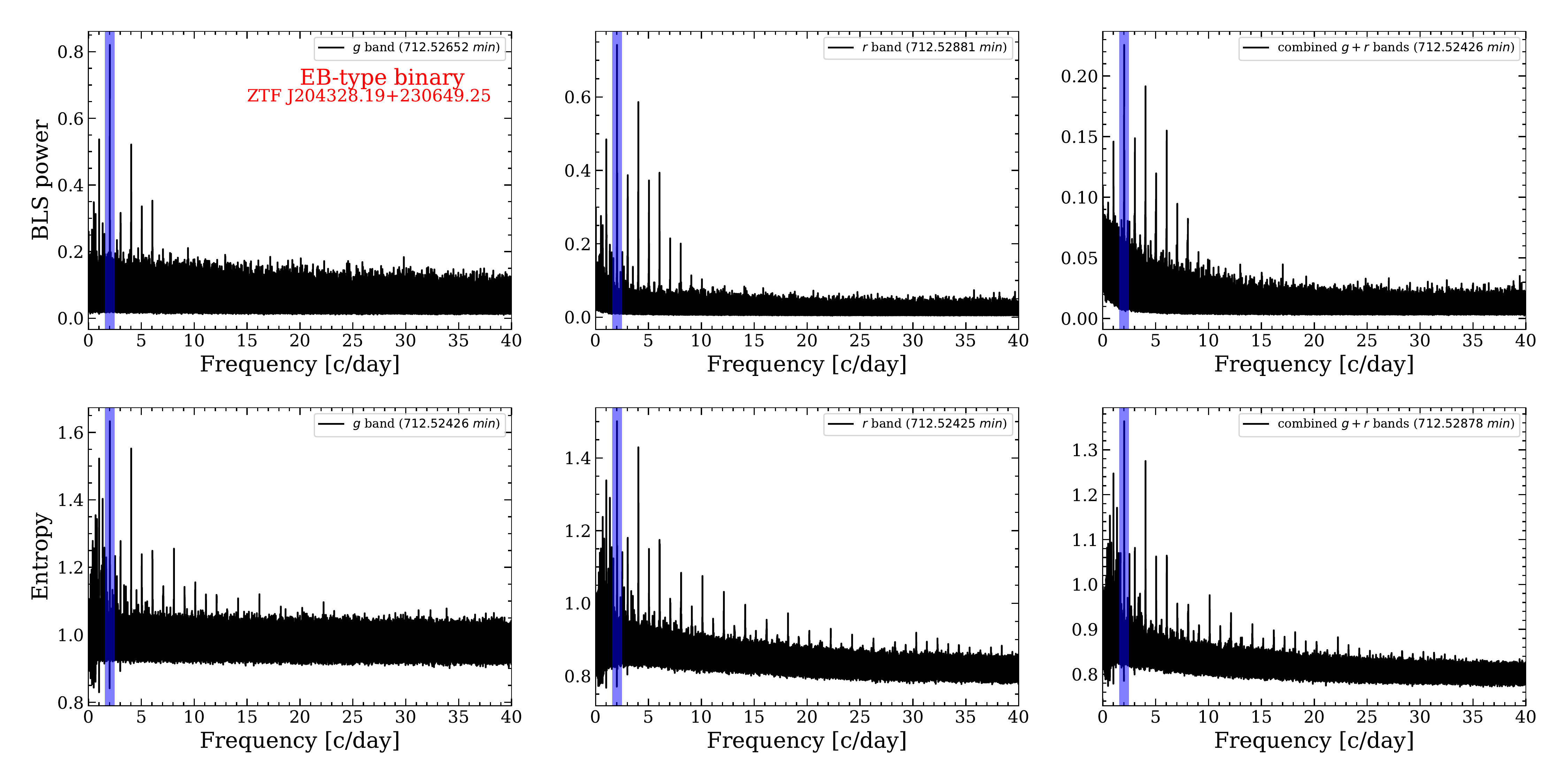}
\includegraphics[width=1.0\textwidth]{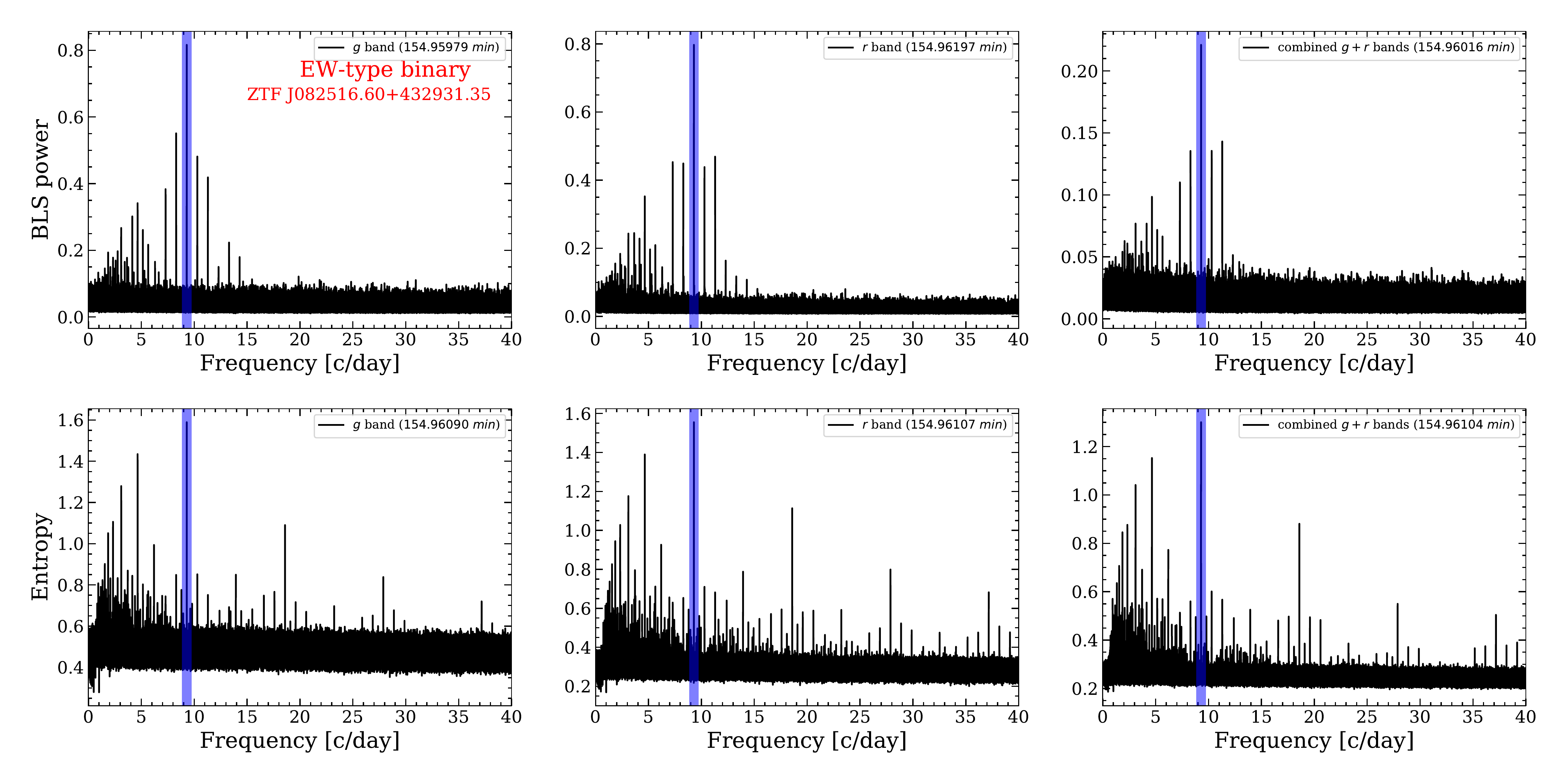}
\caption{Examples of the variability types of light curve shapes caused by the binary eclipse effect in our binary sample. All panels show the frequency spectra of the BLS and CE corresponding to the above (Figure~\ref{fig:LCexample}) variability types of the EB-type binary and EW-type binary.}
\label{fig:BLSCEFS2}
\end{center}
\end{figure*}

\begin{figure*}[htpb]
\begin{center}
\includegraphics[width=1.0\textwidth]{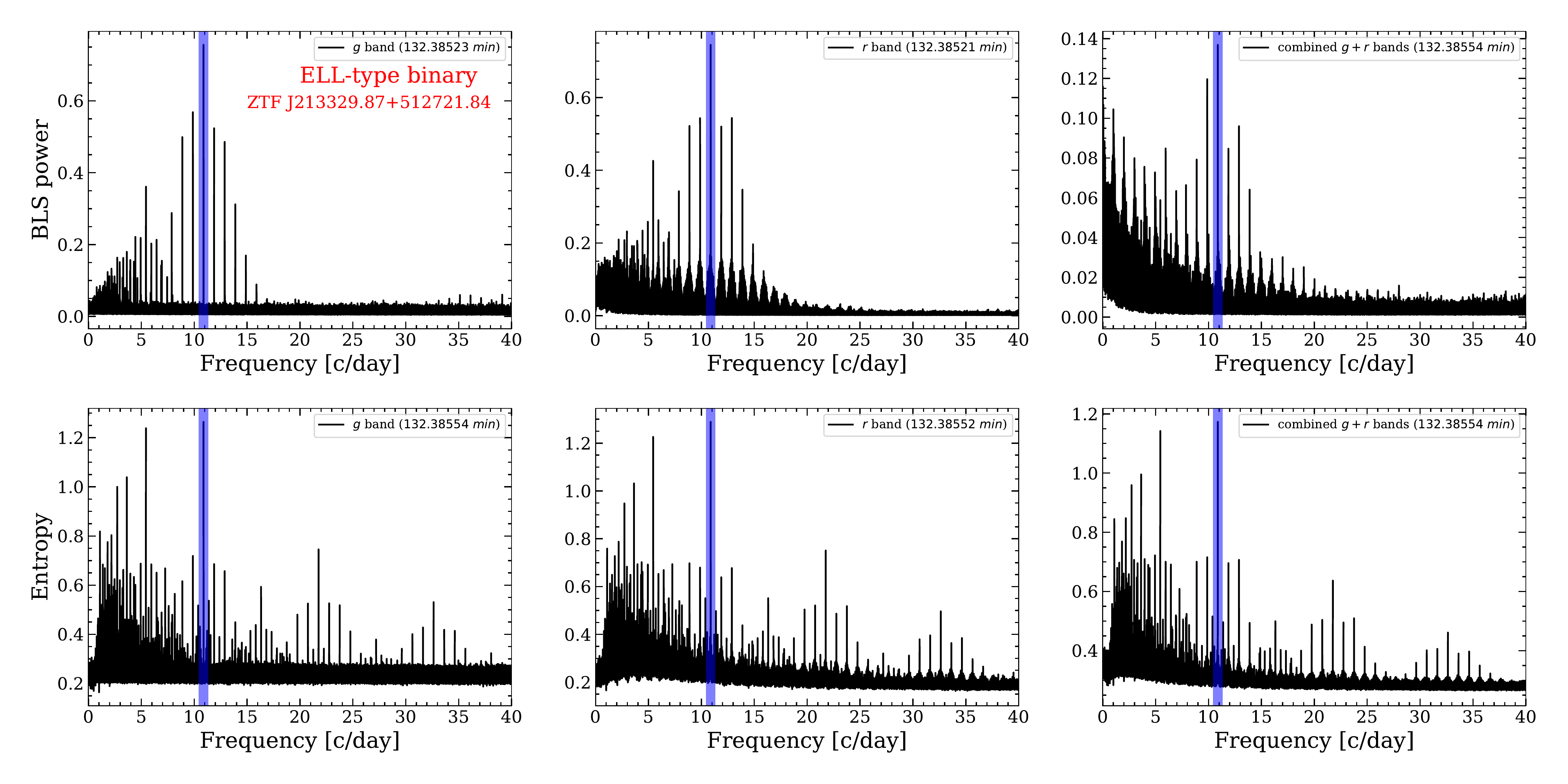}
\includegraphics[width=1.0\textwidth]{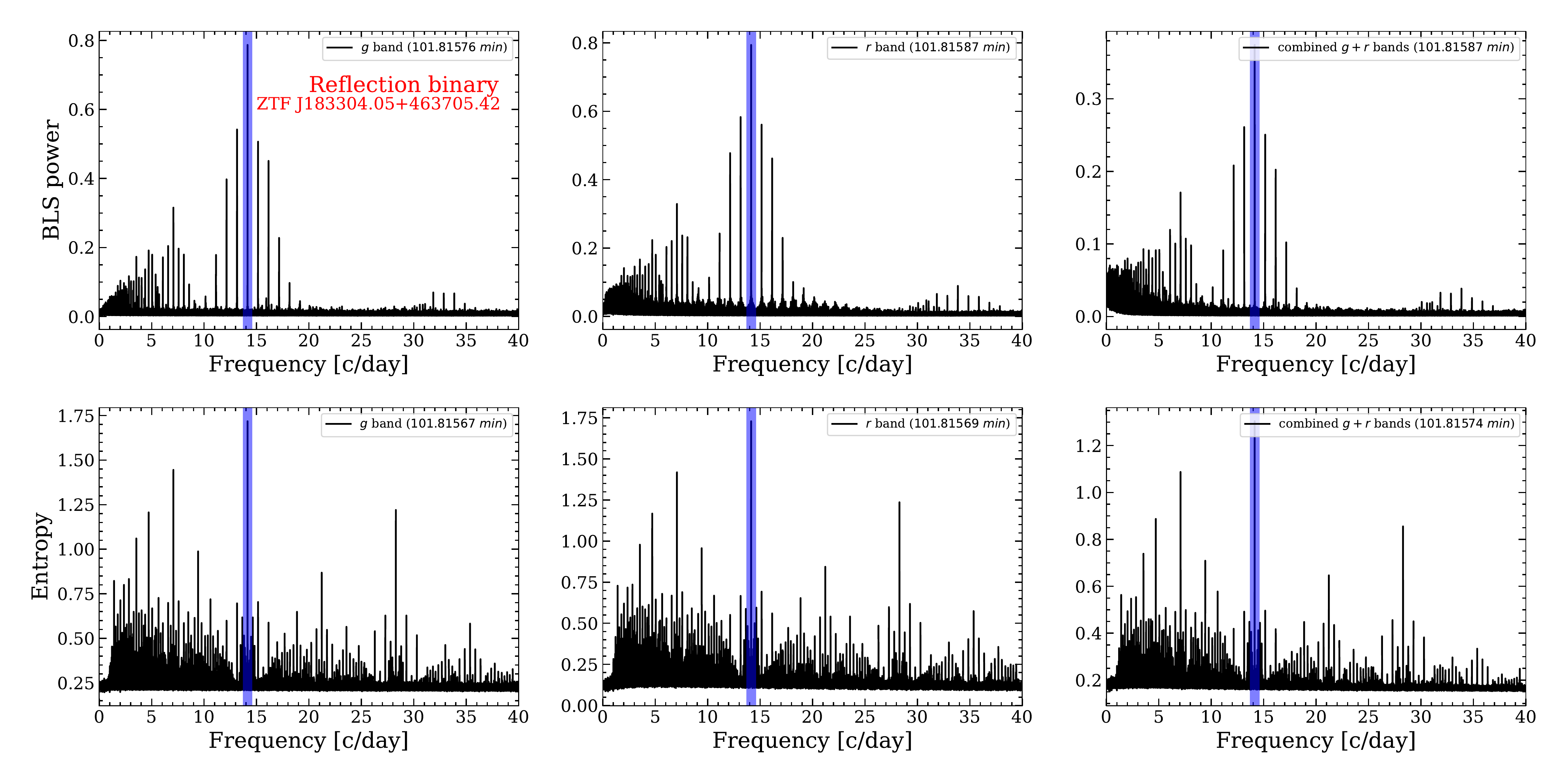}
\caption{Examples of the variability types of light curve shapes caused by the binary eclipse effect in our binary sample. All panels show the frequency spectra of the BLS and CE corresponding to the above (Figure~\ref{fig:LCexample}) variability types of the ELL-type binary and Reflection binary.}
\label{fig:BLSCEFS3}
\end{center}
\end{figure*}

Comparison of the frequency spectrum of BLS and CE with different types of light curves. For further details on the FS, please see Figures \ref{fig:BLSCEFS1} - \ref{fig:BLSCEFS3}.

\section{Light curve-selected Targets} \label{ref:appendixd}
\begin{figure*}[htpb]
\begin{center}
\includegraphics[width=1.0\textwidth]{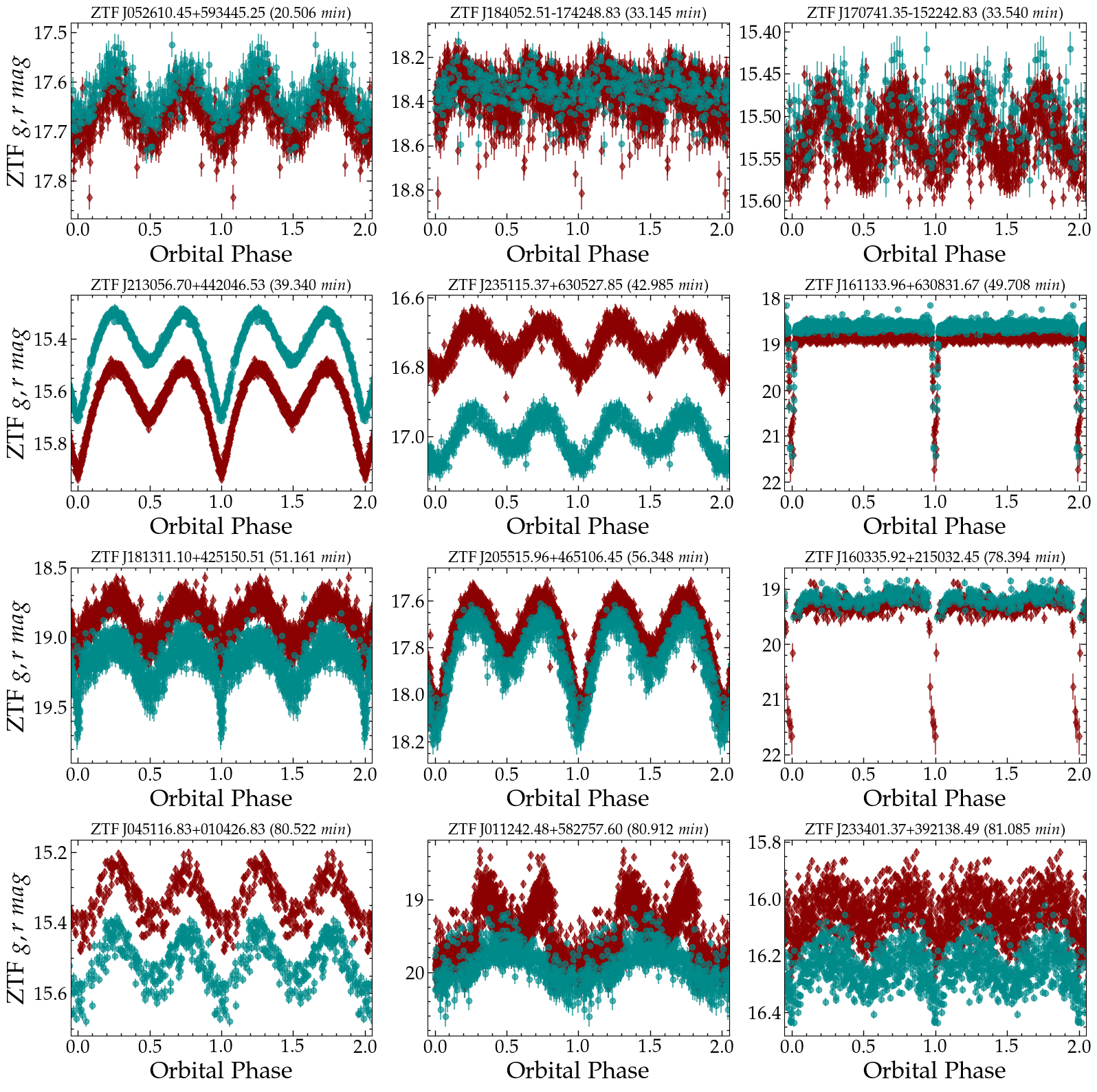}
\caption{Phase-folded ZTF lightcurves of the potential \ac{GW} candidates with an orbital period of less than 100 min, at \ac{GW} amplitude and TianQin/LISA SNRs for these candidates described in Section \ref{subsec:gwsignals}.}
\label{fig:HRPotential1}
\end{center}
\end{figure*}

\begin{figure*}[htpb]
\begin{center}
\includegraphics[width=1.0\textwidth]{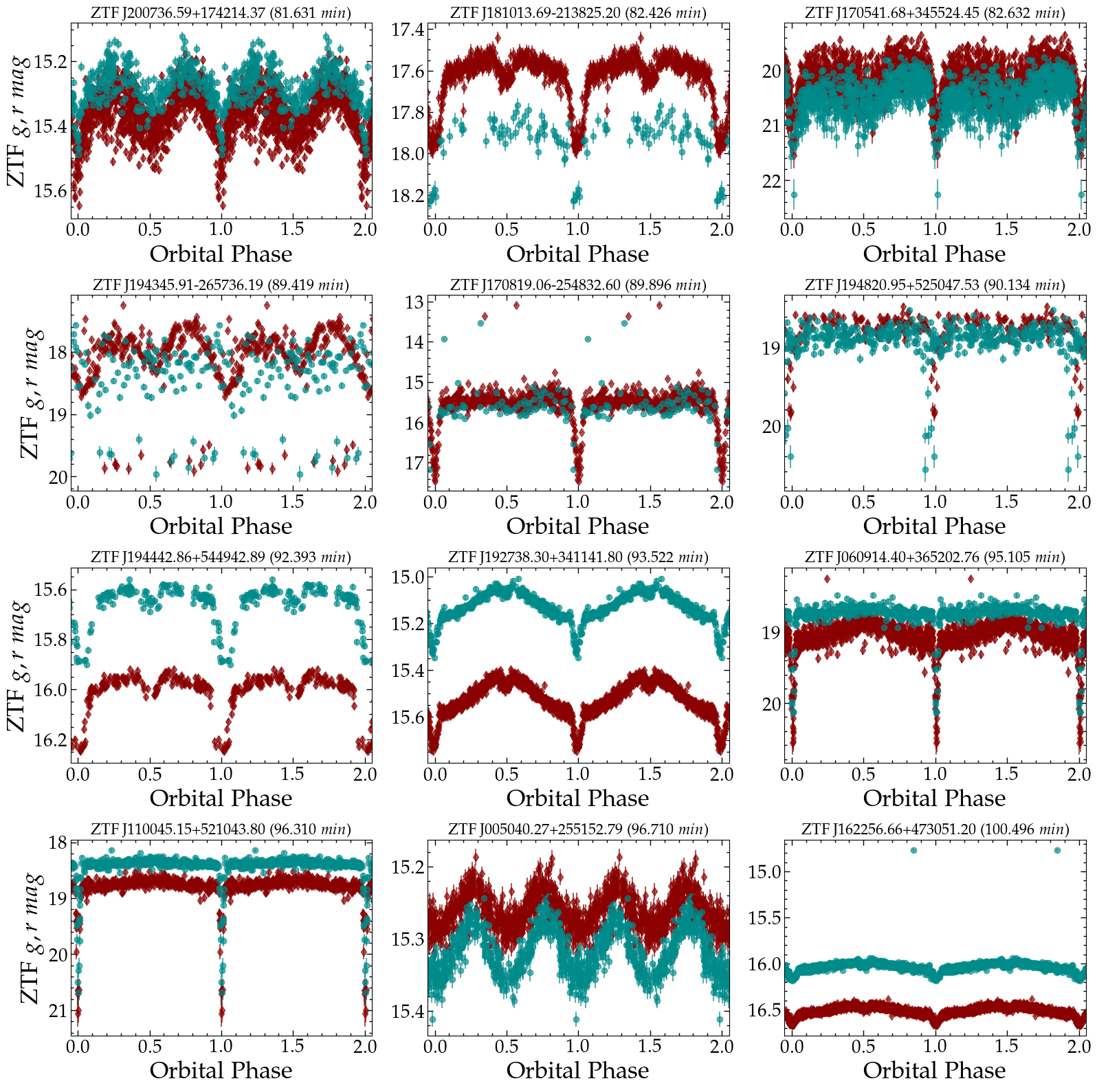}
\caption{Phase-folded ZTF lightcurves of the potential \ac{GW} candidates with an orbital period of less than 100 min, at \ac{GW} amplitude and TianQin/LISA SNRs for these candidates described in Section \ref{subsec:gwsignals}.}
\label{fig:HRPotential2}
\end{center}
\end{figure*}

The phase ZTF ZTF g-band and r-band light curves of potential \ac{GW} source candidates with an orbital period of less than 100 min in the binary sample. For further details on these light curves, please see Figures \ref{fig:HRPotential1} and \ref{fig:HRPotential2} .

\end{CJK*}
\end{document}